\documentclass[traditabstract]{aa}
\usepackage{graphicx}
\usepackage{txfonts}
\usepackage{hyperref}
\usepackage{multirow}
\usepackage{xcolor}
\usepackage{soul} 
\hypersetup{
     colorlinks   = true,
     citecolor    = blue
}
\usepackage{float}
\DeclareUnicodeCharacter{2212}{\textendash}
\begin{document}

    \def \aCO{$\alpha_{\rm CO}\,$}
    \def \aCII{$\alpha_{\rm [CII]}\,$}
    \def \XCI{$\rm X_{CI}\,$}
    \def \gdr{$\delta_{\rm GDR}\,$}

\title{Gas properties as a function of environment in the proto-supercluster Hyperion at $z\sim2.45$}

    \authorrunning{Gururajan et al.}
	\titlerunning{Gas properties in Hyperion}

\author{G. Gururajan
          \inst{1,2,3,4},
          O. Cucciati
          \inst{2},
          B. C. Lemaux
          \inst{5,6},
          M. Talia
          \inst{1,2},
          G. Zamorani
          \inst{2},
          F. Pozzi
          \inst{1,2},
          R. Decarli
          \inst{2},
          B. Forrest
          \inst{7},
          L. Shen
          \inst{8,9},
          G. De Lucia
          \inst{10,4},
          F. Fontanot
          \inst{10,4},
          S. Bardelli
          \inst{2},
          D. C. Baxter
          \inst{11},
          L. P. Cassarà
          \inst{12},
          E. Golden-Marx
          \inst{13},
          D. Sikorski
          \inst{14},
          E. A. Shah
          \inst{5},
          R. R. Gal
          \inst{14},
          M. Giavalisco
          \inst{15},
          F. Giddings
          \inst{14},
          N. P. Hathi
          \inst{16},
          D. Hung
          \inst{6},
          A. M. Koekemoer
          \inst{16},
          V. Le Brun
          \inst{17},
          L. M. Lubin
          \inst{5},
          L. A. M. Tasca
          \inst{17},
          L. Tresse
          \inst{17},
          D. Vergani
          \inst{2},
          \and
          E. Zucca
          \inst{2}
          }

   \institute{
            University of Bologna - Department of Physics and Astronomy “Augusto Righi” (DIFA), Via Gobetti 93/2, I-40129, Bologna, Italy
        \and
            INAF - Osservatorio di Astrofisica e Scienza dello Spazio, Via Gobetti 93/3, I-40129, Bologna, Italy
        \and
            Scuola Internazionale Superiore Studi Avanzati (SISSA), Physics Area, Via Bonomea 265, 34136 Trieste, Italy, \email{ggururaj@sissa.it}
        \and
            IFPU-Institute for Fundamental Physics of the Universe, Via Beirut 2, 34014 Trieste, Italy
        \and
            Department of Physics and Astronomy, University of California Davis, One Shields Avenue, Davis, CA 95616, USA
        \and
            Gemini Observatory, NSF NOIRLab, 670 N. A’ohoku Place, Hilo, HI 96720, USA
        \and
            Department of Physics and Astronomy, University of California Davis, One Shields Avenue, Davis, CA, 95616, USA
        \and
            Department of Physics and Astronomy, Texas A\&M University, College Station, TX, 77843-4242 USA
        \and
            George P.\ and Cynthia Woods Mitchell Institute for Fundamental Physics and Astronomy, Texas A\&M University, College Station, TX, 77843-4242 USA
        \and
            INAF – Astronomical Observatory of Trieste, Via G. B. Tiepolo 11, 34143 Trieste, Italy
        \and
            Department of Astronomy \& Astrophysics, University of California, San Diego, 9500 Gilman Dr, La Jolla, CA 92093, USA
        \and
             INAF- IASF Milano, Via Alfonso Corti 12, 20133, Milano
        \and
            INAF – Osservatorio astronomico di Padova, Vicolo Osservatorio 5, 35122 Padova, Italy
        \and
            Institute for Astronomy, University of Hawai‘i, 2680 Woodlawn Drive, Honolulu, HI 96822, USA
        \and
            Astronomy Department, University of Massachusetts, Amherst, MA 01003, USA
        \and
            Space Telescope Science Institute, 3700 San Martin Drive, Baltimore, MD 21218, USA
        \and
            Aix-Marseille Université, CNRS, CNES, LAM, Marseille, France
            }
         
\abstract { The cosmic star-formation rate density, molecular gas density and the AGN activity of the Universe peak at $z\sim$ 2-3, showing the Universe is most active at this epoch. The nature of the galaxies at these redshifts and their properties as a function of their environment are particularly interesting to understand the mechanisms driving their star-formation and quenching. At $z\sim$ 2.5, a massive ($\sim 4.8 \times 10^{15}\, \rm M_\odot$) proto-supercluster, Hyperion, was identified \citep{Cucciati18}, consisting of 7 groups/peaks and extending over a comoving volume of $60 \times60 \times 150\, \rm Mpc^3$, providing an excellent laboratory to probe the properties and evolution of galaxies as a function of their environments. We use a large compilation of photometric (optical to radio wavelengths, COSMOS2020, COSMOS-Super-deblended, and, A$^3$COSMOS) and spectroscopic (C3VO, \textit{HST}-Hyperion, VUDS, zCOSMOS, DEIMOS10K, MAGAZ3NE) data to assign membership and study the relation between the local environment and the molecular gas mass, the star-formation rate (SFR), gas depletion timescales, and quenching mechanisms. We find that the depletion timescales and the molecular gas fractions decrease and SFR increases in denser environments at the $\sim2\,\sigma$ level, suggesting accelerated evolution in the densest regions of this proto-supercluster resulting from gas stripping, over-consumption, and/or cessation of cold flows. Dedicated observations at sub-millimeter wavelengths enabling further spectroscopic confirmation and better coverage in the sub-millimetric (sub-mm) wavelengths can provide more conclusive results on the environmental implications on gas reservoirs of galaxies in Hyperion. }




\maketitle
\section{Introduction}

The environment at which a galaxy resides plays an important role in its evolution through cosmic time. In the local Universe, galaxies residing at very dense environments, such as in galaxy clusters cores, tend to be massive and quiescent/passive \citep[e.g.][]{Dressler80, Dressler84, Lewis02,Peng12}. Moreover, nearly $25\%$ of the massive galaxies (log(M$_*/ \rm M_\odot) \ga$ 9, up to $z\sim 1.5$) are found to be in clusters or galaxy groups \citep[e.g.][]{Gerke12, Tempel16, Boselli22}. Such massive galaxies in dense environments usually have very low star-formation rate (SFR, \citealt{Gomez03, vonderLinden10, Muzzin12,Tomczak19, Old20}), low molecular gas content \citep[e.g.][]{Fumagalli09, Boselli14}, and tend to show signatures of recent quenching events \citep[e.g.][]{Tran03, Wu14, Lemaux17, Socolovsky18, Owers19, Paccagnella19}. The large number of massive galaxies found in these clusters also suggests that the build-up of stellar mass must have been quite rapid, either due to external processes like mergers or due to a large amount of molecular gas present in them at earlier epochs. In contrast, at redshifts ($z\sim1.5$), some studies have found a population of star-forming, blue galaxies in the less-dense environments of the cluster, whereas the cluster center is still dominated by red, passive galaxies \citep[e.g.][]{Scoville07, McGee11, Muzzin12, Newman14, Strazzullo13, Strazzullo19}. Thus, to understand the evolution of galaxies shaped by their environment, the rapid mass build-up they undergo, and the quenching mechanisms at play, we need to study these overdense structures at higher redshifts.

Proto-clusters of galaxies at high-$z$ ($z \gtrsim 1.5$) are the progenitors of the galaxy clusters seen at the current epochs,  spanning  tens of comoving megaparsecs. By definition, they are to collapse into a galaxy cluster by $z=0$ \citep[e.g.][]{Overzier16}. Proto-clusters of galaxies can be identified by various tracers such as, an excess of star-forming galaxies \citep[e.g.][]{Steidel00, Ouchi05, Lemaux09, Capak11}; Lyman-alpha emitters \citep[LAEs; e.g.][]{Toshikawa14, Toshikawa16, Toshikawa18, Toshikawa20, Dey16} or other line emitters \citep[e.g.][]{Forrest17,Shimakawa18}; Sub-millimeter galaxies \citep[SMGs, e.g.][]{Casey15, Smolcic17a, Greenslade18, Lewis18, Oteo18, Long20, Hill20} or by X-ray emissions \citep[e.g.][]{Fassbender11, Wang16} to name a few. Although, proto-clusters with enhanced quenched fractions are found in the literature \citep[e.g.][]{McConachie22,Ito23,Kakimoto24}, they are often dominated by a population of star-forming galaxies. From a sample of confirmed galaxies residing in environments that spanned more than an order of magnitude in local density between $2\la z \la 5$, including many proto-cluster members, \citet{Lemaux22} analysed 7000 spectroscopically confirmed galaxies and found that the SFR increases at increasing overdensities, the reverse of the trend observed in the local Universe. Additionally, proto-clusters with a dominant population of SMGs also host prodigious SFRs \citep[e.g.][]{Steidel98, Chapman09, Dannerbauer14, Umehata15, Oteo18, Long20, Miller18, Hill20}. Proto-clusters are thus believed to be the drivers of reversal in SFR - overdensity relation at high-$z$.

From the perspective of simulations, proto-clusters also host larger SFRs compared to galaxies in non cluster-like/dense environments (hereafter, `field' galaxies, \citealt{Muldrew18}). In addition to their increased SFRs, proto-clusters also have a larger contribution to the star-formation rate density (SFRD) at a given redshift than the field galaxies. \citet{Chiang17} showed that despite occupying only $\sim 4\%$ in volume at $z\sim 4.5$, the corresponding contribution of proto-clusters can be as high as $\sim 20 \%$ to the cosmic SFRD. Such predictions have also been confirmed at least in one massive proto-cluster at $z\sim4.5$ \citep{Lemaux18, Staab24}. Specifically, \citet{Staab24} found, using the massive proto-cluster \emph{Taralay} as a proxy, that proto-clusters contribute approximately a third of the overall cosmic SFRD at these redshifts.

To understand the nature and star-formation mechanisms of member galaxies in proto-clusters, it is important to characterise their molecular gas content (molecular hydrogen), which is the main fuel of star formation. In addition to knowing the pathways that lead to increased star-formation and rapid mass build-up in the member galaxies of a proto-cluster, it is also important to understand the role of the environment in shaping the accretion of gas onto the gas reservoirs of these galaxies. Galaxies could assemble their stellar mass by accreting a large amount of molecular gas \citep[e.g.][]{Dekel09,Kleiner17,Kretschmer20,Chun20} or may show an increased efficiency in the star-formation for a given amount of molecular gas \citep[e.g.][]{Tacconi13,Dessauges-Zavadsky15,Bethermin15,Aravena16}, yet, the dominant mechanism in proto-clusters leading to rapid mass-growth remains an open question. Due to the lack of a dipole moment of H$_2$ for rotational transitions and as the vibrational transitions of H$_2$ are inaccessible at these redshifts, the molecular content of proto-cluster galaxies cannot be directly observed. Therefore, tracers such as emission lines of CO \citep[e.g.][]{Solomon05,Bolatto13}, [CI] \citep[e.g.][]{Keene85,Papadopoulos04a, Papadopoulos04b} or [CII] \citep[e.g.][]{Zanella18} or the dust continuum emission \citep[e.g.][]{Magdis11, Leroy11,Scoville14} are commonly used as a proxy for the molecular gas. 

In the recent years, several studies have probed the molecular gas content of galaxies in proto-clusters to understand the role of the environment in shaping gas reservoirs. Gas reservoirs are quantified mainly based on the fractional amount of molecular gas present in a given galaxy, i.e. gas fractions, defined either as f$_{\rm gas} = \rm M_{\rm gas}/(M_{\rm gas}+M_{*})$ or as $\mu_{\rm gas} = \rm M_{\rm gas}/M_{\rm *}$ where $\rm M_*$ is the stellar mass of the galaxy. For local clusters, a deficit in gas fractions for cluster galaxies in comparison to the field have been found in the literature \citep[e.g.][]{Fumagalli08, Corbelli12, Morokuma-Matsui21}. Additionally, a decrease in the gas fractions of member galaxies within a cluster from less-dense/filamentary regions to denser regions have also been found for local clusters with CO-based gas mass estimates \citep[e.g.][]{Morokuma-Matsui21, Castignani22} and dust-continuum derived gas mass estimates \citep[e.g.][]{Betti19}. However, some late-type galaxies in the center of clusters are also reported to have higher molecular gas and longer-depletion timescale \citep{Nakanishi06, Mok16}. In the case of intermediate redshift clusters ($1\lesssim z\lesssim 2$), heterogeneous results have been presented regarding their gas reservoirs. While most studies report similar gas fractions in clusters and field galaxies \citep[e.g.][]{Hayashi18, Noble19, Williams22} or increased gas fractions \citep[e.g.][]{Genzel15, Tacconi18}, others report a deficit in gas similar to local clusters \citep{Coogan18, Alberts22}. 

Similarly, in the case of proto-clusters at high-$z$ (i.e. $z>2$), some studies report member galaxies to have similar gas-fractions to the field galaxies at the given redshift \citep[e.g.][]{Lee17, Zavala19, Wang18, Champagne21} while others report galaxies with increased gas fractions \citep[e.g.][]{Tadaki19, GomezGuijarro19}. Such diverse results could arise from differences in the evolutionary stages of different proto-clusters at a given redshift \citep{Muldrew15}, although, most of these studies are limited to small samples of galaxies within a single proto-cluster. Additionally, differences in tracers used such as CO and dust, could lead to inconsistencies in the results as we do not have enough evidence for whether the gas and dust are affected similarly by the environment. Moreover, studies often target the brightest galaxies within a small radius, typically near the proto-cluster center, making it difficult to understand the effects of different environments. Selection of a large galaxy sample in a wide range of overdensities within a proto-cluster could thus help in further understanding the role of environment on the gas reservoir. 

In this work, we present a sample of galaxies associated with the proto-supercluster Hyperion \citep{Cucciati18}. This structure was identified in the COSMOS-field and the density map of Hyperion was constructed using Voronoi Monte Carlo Tesselation by \citet{Cucciati18}. Hyperion spans a volume of $\sim 60 \times 60 \times 150$ comoving Mpc$^3$ in a redshift range of $2.39 \leq z \leq 2.54$ and has an estimated total mass of $\sim 4.8 \times 10^{15}\, \rm M_\odot$. Hyperion consists of seven overdense peaks connected by filamentary regions. These peaks have masses ranging between $0.1-2.7 \times 10^{14} \, \rm M_\odot$, and \citet{cucciati23} have shown that they can be considered bona-fide proto-clusters/proto-groups, based on the study of simulated Hyperion-like structures,  thus making Hyperion a collection of proto-clusters. This makes Hyperion an ideal laboratory not only to study galaxy properties as a function of a wide range of overdensities, but also to study the properties of these individual peaks/proto-groups and explore if they exhibit any differences in their properties amongst the different peaks. 

Individual peaks or regions of Hyperion have been studied previously in the literature \citep{Diener13, Diener15, Casey15,  Chiang14, Chiang15, Lee16,  Wang16, Franck16, Zavala19,Champagne21}. One of the peaks was observed with Atacama large millimeter/sub-millimeter array (ALMA) and Very large array (VLA), targeting the CO emission lines \citep{Wang18,Champagne21}. \citet{Wang18} also observed this peak with XMM-Newton and Chandra and discovered the potential presence of a hot-ICM, and they concluded that the peak could already be a cluster at $z\sim 2.45$. While \citet{Wang18} found that gas fractions decrease and star-formation efficiencies increase for galaxies towards the cluster center (i.e. dense environments), \citet{Champagne21} presented a contrasting view on this proto-cluster. The authors concluded that the X-ray detection was attributed to an inverse-Compton ghost instead of a hot-ICM and find they report the member galaxies to have similar SFEs to the field.

In this paper, we study the member galaxies of Hyperion in the various peaks and filamentary regions. We combine a sample of galaxies with secure spectroscopic redshifts and a sample of galaxies with high quality photometric redshifts to statistically probe the effect of environment on the molecular gas fractions ($\mu_{\rm gas}$), gas depletion timescales ($\tau_{\rm dep} \equiv \,\rm M_{\rm gas}/SFR$) and the SFR within Hyperion as a function of the local-overdensity. This paper is structured as follows. In Sect.\,\ref{sec2}, we present the datasets/catalogues used in this study. In Sect.\,\ref{sec3} we present the spectral energy distribution (SED) fitting to derive the physical properties of our sample and the statistical treatment of our sample of galaxies with photometric redshifts (photo-$z$ sample). Our results are presented in Sect.\,\ref{sec:results} and the discussions of our results are presented in Sect.\,\ref{sec:discussion}. Finally, we summarise the results and present the conclusions of this work in Sect.\,\ref{sec:conclusions}. Throughout this paper, we adopt a $\Lambda$CDM cosmology with $\Omega_m = 0.3$, $\Omega_{\Lambda} = 0.7$ and $H_0 = 70$ km\,s$^{-1}$\,Mpc$^{-1}$ and a \citet{2003PASP..115..763C} initial mass function (IMF).


\section{Sample selection and data availability}\label{sec2}
 
In this paper, we study the member galaxies of the proto-supercluster Hyperion \citep{Cucciati18}. Hyperion falls within the Cosmic Evolution Survey (COSMOS, \citealt{Scoville07}) field, in the range of 149.6 -- 150.52$^{\circ}$ in Right Ascension (RA) and 1.74 -- 2.73$^{\circ}$ in Declination (Dec) and over a redshift range of $2.39\leq z \leq 2.54$. The COSMOS field ($\sim 2$ deg$^2$) has unprecedented depth and multi-wavelength coverage from  X-rays \citep{Elvis09, Civano12, Marchesi16}, Ultraviolet (UV, \citealt{Zamojski07}), Optical \citep{Capak07, Leauthaud07, Taniguchi07, Taniguchi15, Muzzin13}, near-infrared (NIR, \citealt{McCracken10,McCracken12}), mid-infrared (mid-IR, \citealt{Sanders07,LeFloc09}), far-infrared (FIR, \citealt{Lutz11, Oliver12}), sub-millimeter (sub-mm,\citealt{Geach17}), millimeter \citep{Bertoldi07, Aretxaga11} to radio \citep{Schinnerer10, Smolcic17}. We take advantage of the available multiwavelength photometric data in the COSMOS-field to study the properties of galaxies in Hyperion. 

\subsection{Parent photometric catalogue : COSMOS2020}\label{sec:cosmos2020}

Publicly available photometric catalogues in the COSMOS field with photometric coverage from UV to mid-IR (COSMOS2007, \citealt{Capak07}; COSMOS2015, \citealt{Laigle16}; COSMOS2020, \citealt{Weaver22}) have become progressively deeper and larger over the years. In this work, we use the newest available version - COSMOS2020 (henceforth C20) that covers 40 photometric bands, down to a magnitude limit of 25.7 in IRAC channel 1 \citep{Weaver22}. Wavelength coverage of C20 includes near-UV (NUV) and far-UV (FUV) observations from Galaxy Evolution Explorer (GALEX), optical coverage of \textit{g, r, i, z, y} bands with Subaru Hyper Suprime-Cam along with broad, narrow and intermediate bands, Hubble Space Telescope \textit{(HST)} ACS F814W imaging \citep{Koekemoer07}, NIR bands \textit{Y, J, H, K$_s$} from VISTA and mid-IR from \textit{Spitzer-}IRAC channels 1, 2, 3 and 4. In this paper, we adopt the FARMER version, based on the Tractor source profile fitting tool \citep{Lang16}, of the C20 catalogue as our primary photometric catalogue. We apply the galactic extinction corrections from \citet{Schlafly11} as given in the C20 catalogue.

\subsection{Other photometric catalogues}\label{sec:ancillary photo-z cats}

In addition to the UV to mid-IR coverage of the C20 catalogue, observations in the FIR, sub-mm and radio are also available in the COSMOS field. 
The super-deblended photometric catalogue \citep{Jin18} compiles the photometric data from far-infrared to sub-millimetric bands from PACS (100 $\mu m$ and 160 $\mu m$) and SPIRE (250 $\mu$m, 350 $\mu $m, and 500 $\mu $m) on \textit{Herschel} space telescope, SCUBA2 (850 $\mu $m) and AzTEC (1.1 mm) on James Clark Maxwell telescope, and MAMBO (1.2 mm) on IRAM 30 m telescope. The super-deblended catalogue provides point spread function photometric fluxes in the FIR/sub-mm bands by de-blending the sources using deep MIPS-24 $\mu $m images and 1.4\,GHz \citep{Schinnerer10} or 3\,GHz \citep{Smolcic17} VLA observations or $K_s$ band observations from the UltraVISTA catalogues. 

The Automated Mining of the ALMA Archive in the COSMOS Field (A$^3$COSMOS, \citealt{Liu19}) catalogue provides ALMA sub-mm continuum photometric data. The A$^3$COSMOS catalogue provides fluxes of galaxies observed in ALMA bands through targeted and serendipitous detections in the COSMOS-field. In this work, we use the "prior" catalogue of A$^3$COSMOS where the sources are extracted with priors from the UV/Optical catalogues in the COSMOS field.

\subsection{Combining the photometric catalogues in the COSMOS field}\label{sec:combining_photo-z_cats}
The three photometric catalogues put together give us a wavelength coverage from UV to radio, which is essential to recover galaxy properties. To identify the counterparts of the C20 galaxies in the Herschel super-deblended catalogue and in A$^3$COSMOS, we need to identify an angular separation within which a true match can be found. We begin with matching the galaxies in a sufficiently large radius ($\sim 2$ arcsec) and find the local minimum in the distribution of the number of matches between the two catalogues as a function of the angular separation. The local minimum angular separation can be a good indication of the matching radius to employ while matching the catalogues (a similar technique was used in e.g. \citealt{Staab24} and \citealt{Forrest24}). Radii larger than the local minima are likely to be contaminated with false matches. In the case of matches between C20 and Herschel Super-deblended catalogue, a local minimum angular separation was observed at 0.8$^{\prime\prime}$, which was used as the matching radius to find Herschel Super-deblended counterparts of C20 galaxies. Similarly, for A$^3$COSMOS, the local minimum was found at 1$^{\prime\prime}$. With this procedure, our final photometric catalogue consists of C20 galaxies along with their Herschel super-deblended and A$^3$COSMOS counterparts.

\subsection{Spectroscopic catalogues}\label{sec:specz_Catalog}

\subsubsection{Wide-field spectroscopic surveys}

Spectroscopic targets considered in our analysis are primarily derived from the VIsible MultiObject Spectrograph (VIMOS) Ultradeep survey (VUDS, \citealt{Lefevre15}). VUDS is a large spectroscopic survey conducted in three fields - COSMOS, Extended Chandra Deep Field South (ECDFS) and VIMOS-VLT Deep Survey (VVDS), observing galaxies between $z\sim2 - 6$ using the VIMOS spectrograph on Very Large Telescope (VIMOS-VLT). The targets of VUDS are pre-selected to be mostly brighter than $i_{\rm AB} \lesssim 25$ and using the photometric redshift cut from \citet{Capak07}, $z_p + 1\sigma \geq 2.4$ \citep{Lefevre15, Cucciati18}.

\citet{Lefevre15} provide a spectroscopic redshift quality flagging system to label how the redshifts were determined and to quantify the reliability of the spectroscopic redshift measurement. The flags vary from X1, X2, X3, X4 and X9 with probabilities of secure redshifts $\sim$ 50–75$\%$, 75-85$\%$, 95-100$\%$, $100 \%$, and $80\%$ respectively. The prefix `X' is used to denote the nature of the observations where X$= 0$ indicates galaxies, X$= 1$ indicates broad-line AGNs, X$= 2$ indicates serendipitous detections well separated from the target and X$= 3$ denotes serendipitous detections at the position of the main target (mergers or chance alignments). We base our sample selection criteria on the VUDS flagging system including, X2, X3, X4 and X9 galaxies with X$= 0, 2$ or 3 in our analysis, i.e. excluding broad line AGNs. 

The zCOSMOS survey is a spectroscopic redshift survey in the COSMOS-field targeting $0 \lesssim z \lesssim 3 $ using the VIMOS spectrograph on VLT. The zCOSMOS survey consists of two sub-surveys : zCOSMOS Bright \citep{Lilly07,Lilly09} and zCOSMOS Deep spectroscopic survey \citep{Diener13,Diener15}. The bright survey mainly sampled galaxies up to $z\sim1.2$ and $i_{AB} < 22.5$ and the zCOSMOS-Deep survey observed galaxies at a higher redshift $1.4 \leq z \leq 3$. The zCOSMOS sample also has a similar flagging system for the robustness of the spectroscopic redshifts as the VUDS survey. We adopt the same spec-$z$ quality selection criteria for the zCOSMOS galaxies as VUDS. 

Additionally we also include the galaxies from the DEIMOS-10K spectroscopic survey \citep{Hasinger18}. This also samples $\sim 10^4$ objects in the COSMOS-field using the DEep Imaging Multi-Object Spectrograph (DEIMOS, \citealt{Faber03}) instrument on Keck telescopes. Most of the galaxies in this sample are at $z< 2$, but it includes a sub-sample of galaxies at higher-redshifts. We adopt the same spec-$z$ quality flags as VUDS.

\subsubsection{Targeted spectroscopic surveys : C3VO}

The \textit{Charting Cluster construction} with the VUDS and Observations of Redshift Evolution in Large Scale Environments (ORELSE, \citealt{Lubin09}) survey (henceforth, C3VO, \citealt{Lemaux22}) targets candidate overdensities in the COSMOS, ECDFS and CFHTLS-D1 fields using DEIMOS and Multi-Object Spectrometer For Infra-Red Exploration (MOSFIRE, \citealt{McLean08}) on the Keck Telescope \citep{Shen21,Lemaux22,Forrest23,Staab24,Shah24,Forrest24, Hung24}. The target overdensities for C3VO are identified using density maps that were constructed with the available photometry and spectroscopy in these fields. The survey has successfully observed $\sim 2000$ galaxies in the three fields. The catalogue has been described in detail in \citet{Lemaux22,Forrest23,Staab24} and \citet{Forrest24}. The observations and data are presented in \citet{Forrest24}. Observations of Hyperion in the C3VO catalogue were carried out with MOSFIRE masks and are presented in \citet{Forrest24}. The C3VO catalogue initially consisted for DEEP2/ORELSE-style spectroscopic redshift flagging system (see \citealt{Newman13,Lemaux19} for more details), which was then converted to a VUDS-style spectroscopic redshift quality flagging system by \citet{Forrest24}, where the authors included the distinction for serendipitous detections and broad-line AGN. We adopt the flagging system similar to VUDS provided for the C3VO catalogue by \citet{Forrest24} and use galaxies with quality flags X2, X3, X4 or X9 with X$= 0, 2,$ or 3. 

We also include galaxies from the Massive Ancient Galaxies at z$>$3 Near-Infrared Survey (MAGAZ3NE, \citealt{Forrest20}), which specifically targets massive galaxies and their surroundings with Keck/MOSFIRE to study the environmental implications of the properties of massive galaxies in the COSMOS field. Although the survey predominantly targets galaxies at $z\,>3$, a portion of galaxies between $2<z<3$ were spectroscopically confirmed on filler slots and additionally, due to the considerable uncertainties associated with the photo-$z$s at these redshifts, a portion of the targets in this survey were found to fall in the redshift range of $2\leq z \leq 3$. 

\subsubsection{\textit{HST}-Hyperion Survey}

The \textit{HST}-Hyperion survey is a targeted survey with \textit{HST} to obtain spectroscopic confirmation of galaxies in Hyperion (Program ID: HST 16684, PI : Brian Lemaux). The COSMOS field encompassing the Hyperion region was imaged with 25 pointings using \textit{HST/WFC3/F160W} and slitless grism spectroscopy with \textit{HST/WFC3/G141} over 50 orbits of \textit{HST}. We refer the reader to Forrest et al. submitted., for a comprehensive overview on the survey, the target and the data reduction and grism redshift estimation. We use the grism redshifts provided by Forrest et al. submitted. The authors also provide a quality flagging system for the grism redshifts. This system takes into consideration the spectral fitting and the photometric fitting (when available) and assign quality flags $q_f =0 - 5$. We include only those galaxies with grism redshift quality flag $\geq 3$ corresponding to $>\,65 \%$ security in the redshifts. Further details on the assignment of quality flags can be found in Forrest et al. submitted.

Thus, our final spectroscopic catalogue of galaxies in the Hyperion RA-Dec space includes galaxies from VUDS, zCOSMOS, DEIMOS-10K, C3VO, MAGAZ3NE and \textit{HST}-Hyperion. In all the cases except \textit{HST}-Hyperion, the VUDS-style flagging system for the quality of the spec-$z$ is available and hence we uniformly apply the same quality flag requirements to these galaxies. In the case of \textit{HST}-Hyperion targets, the grism-flagging system is different from the VUDS spectroscopic flags and we select galaxies with grism redshift quality flags with reliability $> 65\,\%$ in the determined spectroscopic redshifts. In the case of duplicates in the spectroscopic observations between the various surveys, we give precedence to the highest quality spectroscopic flags. In case the galaxies have the same spectroscopic flags, the order of preference is: C3VO, MAGAZ3NE, VUDS, DEIMOS-10K and zCOSMOS \citep{Forrest24}.


\subsection{Associating spectroscopically confirmed galaxies to their photometric counterparts}\label{sec: spec-z and photo-z match}

To associate the observed photometry to the galaxies in our spectroscopic sample, we find their counterparts in the C20 catalogue using an appropriate angular separation. \citet{Forrest24} matched the spectroscopic galaxies from C3VO, VUDS, zCOSMOS, DEIMOS-10k, MAGAZ3NE with C20-CLASSIC catalogue \footnote{The classic catalogue of COSMOS2020 is based on flux extractions using SExtractor \citep{Bertin96}, where the images are first homogenised to a common point spread function (PSF) and the fluxes are subsequently extracted using circular apertures \citep{Weaver22}.} using various parameters like differences in photometric redshifts, sky coordinates, i and K band magnitudes etc. For more details, we refer to \citet{Forrest24}, where the matching radii/ angular separation for the different spectroscopic surveys to find the C20 counterparts are given. We verify that their criteria hold while matching with FARMER-C20, using the procedure described in Sect.\,\ref{sec:combining_photo-z_cats} and thus adopt them to find the counterparts. We adopt the following matching radii for the galaxies from different spectroscopic surveys : 0.7$^{\prime\prime}$ for galaxies from VUDS, 0.49$^{\prime\prime}$ for zCOSMOS, 1.35$^{\prime\prime}$ for DEIMOS10K and 0.53$^{\prime\prime}$ for C3VO-MOSFIRE.

For $\sim 81.3 \%$ of the spec-$z$ galaxies, we find a counterpart in the FARMER-C20 catalogue. Around $13.3\%$ of the spec-$z$ galaxies do not match with FARMER-C20, but have a counterpart in the CLASSIC-C20. Since $\sim 96\%$ of these CLASSIC matches without a FARMER counterpart fall in the HSC or UltraVISTA masked region of C20 (this mask is applied to remove observations contaminated by bright stars or that fall in the edges of the HSC or Hyper-Suprime cam images, \citealt{Weaver22}), we proceed with the galaxies that are matched to C20 Farmer alone. Additionally, $\sim 5.4\%$ of the spec-$z$ galaxies do not have a match with the FARMER or CLASSIC version of the C20 catalogue. Finally, we found a small number with multiple matches within the Hyperion RA-Dec range, and resolved them manually by assigning each of them to the source with the closest proximity.


\begin{table*}
\centering
\caption{\label{tab:obs_details} Our sample of spectroscopically confirmed Hyperion member galaxies.}
\begin{tabular}{cccccccccc}
\hline
\hline
&&&&&&&&&\\
ID & RA & Dec & Spectroscopic & Redshift &$z$ & $\sigma_\delta$ & SFR  & log(M$_*$)  & log(M$_{\rm gas}$)  \\
&(J2000)&(J2000)&survey&quality flag&&&[M$_\odot$/yr]&[M$_\odot$]&[M$_\odot$] \\
&&&&&&&&&\\
\hline
&&&&&&&&&\\

\textit{HST}2 & 149.94196 & 2.21345 & \textit{HST}-Hyperion & 3$^*$&2.4626 & 2.2 & 87 $\pm$ 8 & 10.18 $\pm$ 0.04 & 10.3 $\pm$ 0.4 \\ 
409222 & 149.93806 & 2.21729 & zCOSMOS & 3.5 &2.4523 & 2.9 & 80 $\pm$ 31 & 10.08 $\pm$ 0.05 & 11.4 $\pm$ 0.1 \\
688810 & 150.34813 & 2.34498 & C3VO-MOSFIRE & 4.0 & 2.4789 & 3.0 & 40 $\pm$ 20 & 10.00 $\pm$ 0.08 & 11.5 $\pm$ 0.2 \\ 
5101211337 & 150.32557 & 2.36018 & VUDS & 3.0 &2.4859 & 3.0 & 39 $\pm$ 4 & 9.94 $\pm$ 0.05 & 11.8 $\pm$ 0.2 \\ 
408290 & 149.74588 & 2.16358 & zCOSMOS & 4.5 &2.4215 & 3.8 & 96 $\pm$ 15 & 10.04 $\pm$ 0.05 & 10.0 $\pm$ 0.2 \\ 
\textit{HST}1 & 150.12676 & 2.36979 & \textit{HST}-Hyperion & 3$^*$ &2.4499 & 3.9 & 28 $\pm$ 11 & 9.97 $\pm$ 0.02 & 10.2 $\pm$ 0.2 \\ 
694682 & 150.35435 & 2.35270 & C3VO-MOSFIRE & 4.0 &2.4562 & 3.9 & 103 $\pm$39 & 10.19 $\pm$ 0.04 & 10.1 $\pm$ 0.1 \\
601715 & 149.98886 & 2.21468 & C3VO-MOSFIRE & 4.0 &2.4262 & 4.4 & 71 $\pm$ 16 & 10.82 $\pm$ 0.03 & 10.34 $\pm$ 0.06 \\ 
409450 & 149.95609 & 2.23012 & zCOSMOS & 3.5 &2.4187 & 4.6 & 53 $\pm$22 & 10.35 $\pm$ 0.07 & 11.6 $\pm$ 0.2 \\
755231 & 150.10534 & 2.44535 & C3VO-MOSFIRE & 4.0 &2.4821 & 5.2 & 90 $\pm$ 11 & 10.56 $\pm$ 0.02 & 9.6 $\pm$ 0.3 \\
617482 & 149.99352 & 2.23855 & C3VO-MOSFIRE & 4.0 &2.4507 & 6.5 & 400 $\pm$ 20 & 10.69 $\pm$ 0.02 & 10.4 $\pm$ 0.1 \\ 
511027238 & 149.99516 & 2.23734 & VUDS & 3.0 &2.4485 & 6.5 & 117 $\pm$ 8 & 10.15 $\pm$ 0.03 & 10.2 $\pm$ 0.2 \\ 
511234645 & 150.11432 & 2.36992 & VUDS & 2.0 &2.4677 & 7.0 & 117 $\pm$ 38 & 10.42 $\pm$ 0.04 & 10.1$\pm$ 0.2 \\
511233051 & 150.07718 & 2.38048 & VUDS & 4.0 & 2.4665 & 7.6 & 131 $\pm$ 17 & 10.97 $\pm$ 0.08 & 10.4 $\pm$0.1 \\
\hline
\end{tabular}
\tablefoot{The spectroscopic/grism redshift ($z$), the spectroscopic redshift quality flag, the right ascension (RA) and declination (Dec) of our spectroscopic sample are tabulated along with their physical properties. In the case of the galaxies from \textit{HST-}Hyperion (indicated by $*$), the grism redshift quality flag is tabulated. The IDs are from the corresponding spectroscopic survey from which the redshifts are obtained \citep{Forrest24}. The $\sigma_\delta$ corresponds to the overdensity level associated to the galaxy in the Hyperion VMC density field \citep{Cucciati18}. The SFR and M$_*$ presented here are obtained by fitting the available photometry from UV to radio for the galaxy using \texttt{CIGALE} (see Sect.\,\ref{sec:sed_fitting_CIGALE}). The gas masses are computed from the dust mass obtained with fitting the FIR/sub-mm SED of the galaxy with a modified black body and assuming \gdr $=100$ (see Sect.\,\ref{sec:sed_fitting_MBB}).}
\end{table*}


\subsection{Associating spec-$z$ galaxies to Hyperion}\label{Sec:spec-z galaxies to Hyperion}

The density mapping of Hyperion was computed by \citet{Cucciati18}, using a Voronoi Monte Carlo (VMC) tessellation mapping technique. We invite the reader to refer to \citet{Cucciati18} for more details on the VMC mapping of Hyperion. To summarise, the authors estimate the local density around galaxies based on spectroscopic galaxies from VUDS+zCOSMOS, and COSMOS2015 photometric redshifts for galaxies without a spec-$z$. First, the 2D local density is computed in overlapping redshift slices, using the Voronoi tessellation, while taking into account the photo-$z$ uncertainties along the line of sight via a Monte Carlo approach. As a second step, the density field in each slice is mapped onto a regular RA-Dec grid. The over-density ($\delta$) field is then computed, based on the local density and its mean value. The distribution of log$(1+\delta)$ of all pixels in each slice is then fit with a Gaussian, and the environment at each pixel is finally parameterised based on the parameter $\sigma_\delta$, which is the number of $\sigma$, based on the Gaussian fit, corresponding to each log$(1+\delta)$ value. The slices are then piled-up in redshift so to build the third dimension. The result is a data cube, where in each voxel the local environment is parameterised in terms of  $\sigma_\delta$. Hyperion is identified as the contiguous volume of the data cube with $\sigma_\delta \geq 2$. The authors also define regions with $\sigma_\delta \geq 5$ as the `peaks of Hyperion' and have identified 7 of such overdensity peaks. Overall, Hyperion spans a redshift range between $2.39 \leq z \leq 2.54$.

In this work, we use $\sigma_\delta$ as the parameter to define the environment of galaxies within Hyperion. We associate the RA-Dec-redshift of a given spectroscopically confirmed galaxy to the corresponding voxel in the VMC map, and we assign to that galaxy the corresponding $\sigma_{\delta}$ value. When the galaxy is associated to a voxel with $\sigma_\delta \geq 2$ within the Hyperion contours, the galaxy is deemed as a Hyperion member. All other spec-$z$ galaxies are not used in our analysis.

\subsection{Combined Optical/NIR and FIR sample}\label{sec:sample selection for SED fitting}

The main objective of this paper is to understand the role of the environment in driving the evolution and properties of galaxies. Here, we focus specifically on the environmental effects on the molecular gas content of galaxies. To probe the molecular gas content, we rely on the dust as a tracer of the molecular gas mass \citep[e.g.][]{Magdis11, Leroy11,Eales12,Scoville14, Scoville16}. 

The dust mass, the dust luminosity and the obscured star-formation in a galaxy can be estimated by fitting its spectral energy distribution, specifically in its FIR wavelengths. Thus, for our further analysis, we select the sub-sample of galaxies with $\geq 2\,\sigma$ detection in at least one FIR/sub-mm bands (100 $\mu m$ -- 2.6 mm). Our final sample comprises of 14 spectroscopic Hyperion member galaxies (2 from \textit{HST}-Hyperion and 12 from ground-based spec-$z$ surveys). In the case of galaxies without a secure spectroscopic/grism redshift and only with a photometric redshift, i.e. the photo-$z$ sample, we apply the following criteria: the galaxies must have a $\geq 2\,\sigma$ detection in at least one FIR/sub-mm band, must have their RA-Dec coordinates in the projected Hyperion RA-Dec area, and a non-zero probability on their integrated photo-$z$ to fall within the Hyperion redshift range ($2.39\leq z \leq 2.54$). For this, we compute the integrated probability of the photo-$z$ to be within $2.39\leq z \leq 2.54$ from the probability distribution of the photometric redshifts from LePhare from \citet{Weaver22}. This gives us 616 photo-$z$ galaxies which are potential Hyperion member-candidates (see Sect.\,\ref{Sec:spec-z galaxies to Hyperion} for associating local overdensities to galaxies). Applying these selection criteria in the FIR and photo-$z$ probability, we find that about $\sim 51 \%$ of our photo-$z$ sample and $\sim 57\,\%$ of our spectroscopic galaxies have at least one FIR band detected with $\geq 3\,\sigma$ and $\sim 41\%$ of our photo-$z$ sample and $\sim 78\%$ of our spec-$z$ sample have multiple bands detected at $\geq 2\,\sigma$ in the FIR in addition to detections in the UV, optical bands and radio bands. Our final spec-$z$ sample is presented in Table\,\ref{tab:obs_details}.

\section{SED fitting}\label{sec3}

The multi-wavelength spectral energy distribution of a galaxy encodes the emission from various components of the galaxy such as the stars, gas and dust emission. Modelling the multi-wavelength SED of a galaxy could thus help us constrain its physical properties such as stellar mass, star-formation rates and dust mass. In this work, we use two SED-fitting methods to constrain the physical properties of the member-galaxies of Hyperion.

\subsection{SED Fitting using \texttt{CIGALE}}\label{sec:sed_fitting_CIGALE}

We estimate the stellar mass (M$_*$) and the total star formation rates (SFR) of our sample using the SED fitting tool \texttt{CIGALE} \citep{Boquien19}. \texttt{CIGALE} employs the energy balance principle, that accounts for the dust absorption in UV/optical and the re-emission in the IR wavelengths. \texttt{CIGALE} models the components of galaxies, such as stars, dust, nebular emission of galaxies, and compares the models to observations and subsequently estimates physical properties such as stellar mass and star-formation rate in a Bayesian approach. 

We use the photometric fluxes from UV to NIR from the C20 catalogue, the FIR and radio (3\,GHz) fluxes from the \textit{Herschel}-super-deblended catalogue, and the sub-mm fluxes from the A$^3$COSMOS catalogue. To perform the SED fitting using \texttt{CIGALE}, we select models that describe the star-formation history (SFH), single stellar population (SSP), nebular emission, dust attenuation, dust emission and emission of the radio component. In Table\,\ref{tab:sed_params}, we describe the various parameters of each of these models used for our SED fitting. 

We use a delayed star-formation template with a burst component to describe the SFHs of our galaxies. In this module \citep{Boquien19}, the SFR is described as :

\begin{equation*}
    \rm SFR (\it t) \propto \frac{\it t}{\tau_{\rm main}^2} \,\times e^{- \it t/\tau_{\rm main}}\,\, \rm for\,\,  0 \leq \it t \leq \it t_0,
\end{equation*}

where $t_0$ is the time of onset of star-formation and $\tau_{\rm main}$ is the e-folding time of the stellar population. The module also contains some additional parameters such as the age of the main population, folding time and age of the burst and the amplitude of the burst $f_{\rm burst}$. The amplitude is described as a mass ratio between the stellar mass in the burst and the total stellar mass. To describe the SSP, we use the \citet{BruzualCharlot03} SSP models with a \citet{2003PASP..115..763C} IMF for solar and sub-solar metallicities. To model the nebular component of the galaxy, we use solar and sub-solar metallities and an ionisation parameter of $-3.0 \,\rm and \, -2.0$ for the nebular region.

\begin{table*}[]
\centering
\caption{Parameters used for SED fitting with CIGALE}\label{tab:sed_params}
\begin{tabular}{ccc}

\hline
\hline
&&\\
Modules & Parameters & Description\\
\hline
&&\\

SFH& sfh delayed& delayed Star Formation History with optional exponential burst  \\
tau$\textunderscore$main (Myr)& 1000,1500,2000,2500& e-folding time of the main stellar population mode\\
age$\textunderscore$main (Myr) & 1000,2000&Age of the main stellar population in the galaxy \\
tau$\textunderscore$burst (Myr) & 100 &e-folding time of the late starburst population model \\
f$\textunderscore$burst & 0.0,0.05&Mass fraction of the late burst population\\
SSP& \citet{BruzualCharlot03}& Single stellar population \\
IMF&\citet{2003PASP..115..763C}& Initial mass function\\
Z&0.2,1& Metallicity, Z$_\odot = 0.02$\\
log(U)&-3.0,-2.0&Ionisation parameter of the nebular component\\
Z$_{\rm gas}$ & 0.6,1& Gas metallicity, Z$_\odot = 0.02$\\
Dust attenuation&\citet{CharlotFall00}&\\
Av$\textunderscore$ISM & 0.3,1.0,1.7&V-band attenuation in the interstellar medium\\
$\mu$&0.3,0.44&$\rm A_{\rm V,ISM}/ (A_{\rm V,BC}+A_{\rm V,ISM})$\\
Dust emission & \citet{Draine14}&\\
q$_{\rm pah}$ & 1.12,2.5,3.9& Mass fraction of PAH\\
U$_{\rm min}$&1.0,5.0,10.0& Minimum radiation field\\
$\alpha$&1.5,2.0,2.5,3.0&Power-law slope\\
$\gamma$&0.02,0.1& Fraction illuminated from U$_{\rm min}$ to U$_{\rm max}$\\
q$_{\rm IR}$ & 2.19,2.58&The value of the FIR/radio correlation coefficient for star formation.\\
\hline\\

\end{tabular}
\tablefoot{We refer the reader to \citet{Boquien19} for a detailed description of each of the SED fitting parameter.}
\end{table*}

Dust is an important component in a galaxy. The light from stars, stellar-remnants, and nebular region is absorbed and re-emitted by the dust present in a galaxy. This absorption and re-emission of the dust needs to be accounted in our dust models to properly estimate the total SFR in a galaxy. We use the \citet{CharlotFall00} recipe for the dust attenuation and the \citet{Draine14} for dust emission. In the case of dust attenuation models, \citet{CharlotFall00} take into account the contribution by the young stars ($<10$ Myr) and the older stellar population separately. In the case of the young stellar population, their emission is attenuated both by the dust in the birth cloud and in the ISM, while the older stars are attenuated by only the dust in the ISM. For both these cases, the attenuation is modelled as a power law normalised to the attenuation in the $V$-band,
\begin{equation*}
    A_V^{\rm ISM} = A_\lambda^{\rm ISM} (\lambda_V/\lambda)^\delta \,\,;
    \, \, \mu = A_V^{\rm ISM}/(A_V^{\rm ISM}+A_V^{\rm BC}),
\end{equation*}
where $\delta$ is the slope of the power law ($-0.7$ for ISM and $-1.3$ for birth clouds, \citealt{Boquien19}). The total attenuation is estimated using the $V$-band attenuation through the ISM ($A_V^{\rm ISM}$) and the ratio ($\mu$) between the attenuation of the old stars (older than 10 Myr) to the young stars (younger than 10 Myr). 

To model the dust emission in our sample, we use the \citet{Draine14} module in \texttt{CIGALE}. These templates constrain the PAH emission and model the dust continuum emission considering the dust heated from the general stellar population and from the star-forming regions separately. The \citet{Draine14} templates have the following updates in comparison to the \citet{DraineLi07}: increased range of radiation field intensities and PAH mass fractions and the power-law slope ($\alpha$) that fits dust emission linked to star-formation is now a free parameter. We also use a radio component module of \texttt{CIGALE} that models the radio-synchroton emission since we have the 3\,GHz band observed for most of our galaxies. We choose the value for IR-radio correlation from \citet{Delhaize17} along with the power-law slope for the star-forming galaxies.

\subsection{Constraining dust mass using Modified Black Body fit}\label{sec:sed_fitting_MBB}

The thermal emission of the dust component of galaxies can be described by a single temperature Modified Black Body (MBB, \citealt{Casey12}):
\begin{equation*}
    S_\nu \propto (1-e^{-\tau_\nu})B_\nu(T),
\end{equation*}
where $B_\nu(T)$ is the black body radiation at temperature $T$ and $\tau_\nu$ is the frequency dependent optical depth. 

In the case of an optically thin dust scenario, the thermal emission can be linked to the dust mass and to the dust emissivity index \citep[e.g.][]{Gilli14, Pozzi20} as :

\begin{equation}\label{SED_fit_eq}
    S_\nu = \rm M_{\rm dust}(1+z)D_L^{-2}\kappa_0(\nu/\nu_0)^\beta \, \textit{B}_\nu(T_{\rm dust}).
\end{equation}

To estimate the dust masses of our sample, we fit the photometric points in the wavelength range 100 $\mu $m - 2.5 mm with the simplified modified Black body (MBB) function described in Eq.\,\ref{SED_fit_eq}. As $\sim 45 \%$ of our photometric sample are dominated by upperlimits, we use this simplistic approach to avoid over fitting the dust emission with complicated models that contain many parameters. We adopt the following value of $\kappa_0 = 0.486 \rm \,cm^2g^{-1}$ at $\nu_0 = 353$ GHz ($\lambda_0 = 850\,\mu$m) from \citet{Draine14}. We also fix $\beta=2$ \citep{Magnelli12} and T$_{\rm dust} = 35$ K for our galaxies, thus leaving the SED normalisations as the free parameter \citep[e.g.][]{Magnelli14, Pozzi20, Traina24}.

For every galaxy, we determine the best-fit SED normalisation and the associated dust mass using a Bayesian Monte Carlo Markov chain approach, with a $\chi^2$ likelihood function to determine the posterior probability distribution of the dust mass. The dust mass is sampled in a uniform prior in log space, between 10$^6$ - 10$^{12}$ M$_\odot$. We use the \textit{python} package \texttt{EMCEE} \citep{Foreman-mackey13} to run the MCMC. We estimate our $\chi^2$ likelihood function using the formulation proposed by \citet{Sawicki12} and \citet{Boquien19} :
\begin{equation}
    \chi^2 = \sum_{n=1}^{n=N}\left ( \frac{f_n-m_n}{\sigma_n} \right )^2 - 2\sum_{0}^{n=N} ln\left \{ \frac{1}{2}\left [ 1 +  \textup{erf}\left ( \frac{f_{n,ulim}-m_n}{\sqrt2\sigma_n} \right ) \right ] \right \},
\end{equation}
where $N$ is the number of photometric bands for a given galaxy, $f_n$ is the flux if it is a detection ($\geq 2\sigma$) and $f_{n,ulim}$ if it is an upper-limit; $m_n$ is the model flux at the photometric band and $\sigma_n$ is the observed error on the fluxes. We then estimate our best-fit model by maximising the ln(likelihood) from running the MCMC with 250 walkers and 1000 chain steps. The best-fit dust mass and the $850\mu$m rest-frame fluxes from the best-fit model are then used in further analysis.

\subsection{Treatment of spec-$z$ and photo-$z$ galaxies}\label{sec:photo-z treatment}

To estimate the physical properties of the spec-$z$ members of Hyperion, we perform the two SED fitting (\texttt{CIGALE} and MBB) at the given spectroscopic redshift of the members. We obtain the total SFR and the stellar mass from the \texttt{CIGALE} SED fitting and the dust mass and rest-frame 850$\mu$m luminosities from the MBB SED fitting. 

In the case of the photo-$z$ members, we estimate the overdensity associated to each of our galaxies from the Hyperion 3-D cube using their RA-Dec-Redshift. In the case of the galaxies with spectroscopic/grism redshift, we can assign a value of the local overdensity directly using the secure spec-$z$. In the case of the photometric sample, due to the lack of robust spectroscopic redshift, we use a statistical approach to estimate their environmental and physical properties. 

The photometric redshift and the photo-$z$ probability distribution function (PDF) of these galaxies are derived in C20 using both LePhare \citep{Arnouts2002, Ilbert06} and EAZY \citep{Brammer08}. In our analysis, we use the LePhare photo-$z$ PDF. For every galaxy in our photo-$z$ sample, we draw an object 1000 times from the C20-LePhare photo-z PDF.

At each value of z$_i$ that falls within $2.39 \leq z \leq 2.54$ taken from the \textit{i-}th realisation, we associate an overdensity value by extracting the $\sigma_\delta$ from the Hyperion cube at the RA, Dec of the galaxy and the redshift value z$_i$ of the realisation (Sect.\,\ref{Sec:spec-z galaxies to Hyperion}). For a given realisation $i$, if the galaxy is associated to a voxel with $\sigma_\delta \geq 2$ within Hyperion, we flag the galaxy to be a Hyperion-member in the realisation $i$. Over all realisations, the median number of photo-z objects that are classified as Hyperion members per realisation was found to be 17, a sample comparable to that of our spectral member sample (i.e. 14 galaxies).

To estimate the physical properties of the photo-$z$ galaxies, we perform the \texttt{CIGALE} and MBB SED fitting at three redshifts, $z=$ 2.39, 2.46 and $2.54$ corresponding to the redshift range spanned by Hyperion, $z = 2.39-2.54$ and thus obtain 3 values of SFR, $\rm M_{*}$ and $\rm M_{\rm dust}$ estimates with their associated errors. This gives us a redshift-parameter (SFR, $\rm M_{*}$ and $\rm M_{\rm dust}$) relation for each of our galaxies. When a galaxy is flagged as a Hyperion member in a given realisation $i$, we interpolate the SFR[$i$], error-SFR[$i$], M$_*$[$i$], error-M$_*$[$i$], M$_{\rm dust}$[$i$] and error-M$_{\rm dust}$[$i$] of the galaxy at $z[i]$. This gives us 1000 realisations of redshifts, and a varying number of $\sigma_\delta$, SFRs, $\rm M_{*}$ and $\rm M_{\rm dust}$ and their associated errors for each of the galaxies, which are used in our further analysis.


\section{Results}\label{sec:results}

In this section, we study the environmental dependence of the properties of the gas reservoirs hosted by Hyperion member galaxies. To estimate the molecular gas mass of our galaxies we use their dust content as a tracer of molecular gas. It can be estimated from the dust mass of a galaxy assuming a gas-to-dust ratio $\delta_{GDR}$ ($\rm M_{\rm mol} = \delta_{\rm GDR} \times M_{\rm dust}$; \citealt{Magdis11,Leroy11,Scoville16}). We adopt the commonly used value for normal star-forming galaxies with solar-like metallicities $\delta_{\rm GDR} = 100$ for our sample \citep[e.g.][]{Leroy11, Magdis12, Remyruyer14, Valentino18, Dunne21}. In addition, we estimate the 850$\mu m$ rest-frame fluxes for our galaxies and estimate the gas mass using the L$_{850}\,\mu$m - M$_{\rm mol}$ relation from \citet{Scoville16}. The gas masses computed with the dust mass from the MBB equation (Eq.\,\ref{SED_fit_eq}) and derived from the 850$\mu m$ fluxes have a broad agreement ($\sim 1\,\sigma$). Thus, we use the dust-mass derived molecular gas mass with $\delta_{\rm GDR}$ in our analysis. 

From the molecular gas mass of a galaxy, we can estimate the depletion timescales $\tau_{\rm dep} \equiv \rm M_{\rm mol}/SFR$ and the gas fractions $\rm \mu_{\rm gas} \equiv M_{\rm mol}/M_{*}$ of the galaxy. The molecular gas depletion timescale can be used to understand the timescales at which the molecular gas content of a galaxy can be depleted due to the star-formation. In other words, this factor is an inverse of the efficiency of star-formation (SFE $\equiv$ SFR/M$_{\rm gas}$). The molecular gas fraction quantifies the availability/lack of the molecular gas in a galaxy and its distribution as a function of the environment. In this section, we study evolution of the depletion timescale (or SFE) and the gas fractions of a galaxy as a function of its local environment to probe the environmental effects in evolution of galaxies in Hyperion. 

\begin{figure*}
\centering
\begin{tabular}{cc}
\includegraphics[width=9cm]{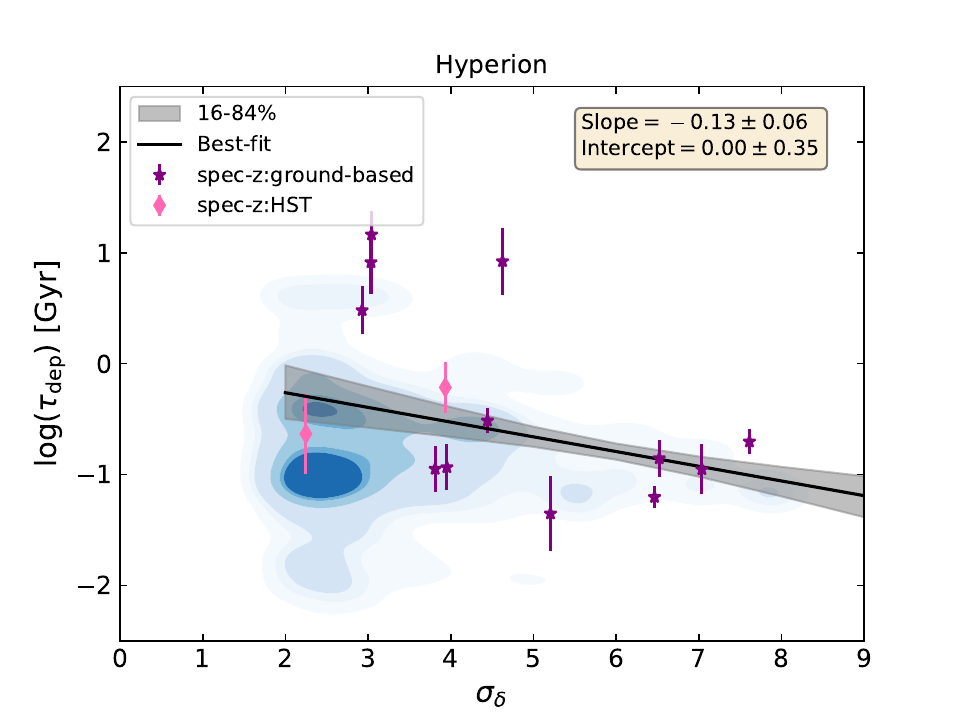}&  
\includegraphics[width=9cm]{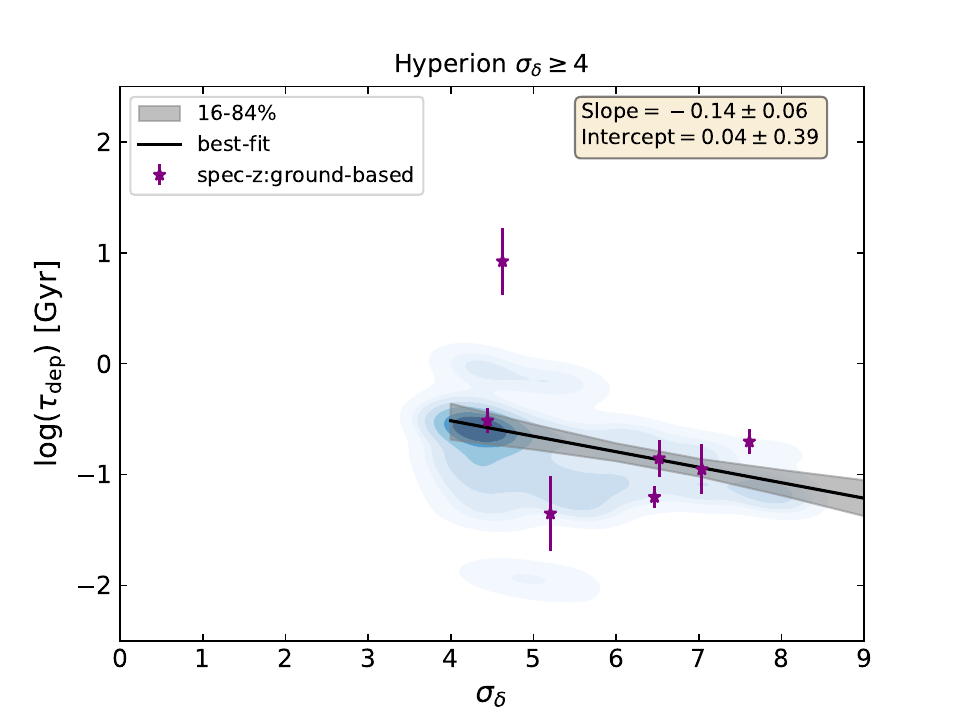} \\
\end{tabular}

\caption{\label{fig:tdep_sigma_slopes}Molecular gas depletion timescale as a function of the environment, quantified by $\sigma_{\rm \delta}$ – the number of standard deviations from the mean in the log($1+\delta$) distribution of local overdensity. In both the panels, the blue-contours represent the distribution of the photo-$z$ galaxies from their 1000 realisations. The blue-contours are the 0.05, 0.16, 0.25, 0.5, 0.75, 0.84 and, 0.9 quantile levels of photo-$z$ galaxies distribution from their 1000 realisations. The purple stars are the spec-$z$ Hyperion members, the pink diamonds represent the spec-$z$ galaxies from \textit{HST}-Hyperion sample. The best fit relation and the 1-$\sigma$ error are shown as the black-solid line and the grey shaded regions respectively. The best-fit slopes and intercepts corresponding to each plot are written in the inset-yellow panel. The errors on the slopes and intercepts are the standard deviation of the slopes and intercepts of the 1000 relations. The left panel shows the $\tau_{\rm dep}- \sigma_\delta$ relation for photo-$z$ and spec-$z$ Hyperion galaxies. The right panel shows the $\tau_{\rm dep}- \sigma_\delta$ relation of Hyperion member galaxies with $\sigma_\delta \geq 4$. } 

\end{figure*}


Our final sample of Hyperion galaxies consists of both galaxies with a reliable spec-$z$ as well as galaxies for which we rely on their photo-$z$s due to the absence of reliable spectroscopic information. To study the environmental dependence of $\tau_{\rm dep},\, \rm \mu_{\rm gas}$ and SFR in Hyperion, we need to use an approach that can allow us to combine the two samples. Hence we make 1000 realisations of $\tau_{\rm dep},\, \rm \mu_{\rm gas}$ and SFR and $\sigma_{\delta}$ relation, where, 
\begin{itemize}
    \item In every realisation, we include the spectroscopic sample (14 galaxies).
    \item In a given realisation \textit{i}, we include only those photo-$z$ galaxies which can be classified as Hyperion members in realisation \textit{i} (see Sect.\,\ref{sec:photo-z treatment} for the analysis on photo-$z$ galaxies).  
    \item For both, spec-$z$ and photo-$z$ galaxies, we add an additional random error drawn from a gaussian distribution whose sigma/width is determined from the uncertainty of each quantity for each galaxy, for every realisation to compensate for the inclusion of spec-$z$ galaxies in every realisation. 
\end{itemize}

Thus, for every $\tau_{\rm dep},\, \mu_{\rm gas},$ and SFR versus $\sigma_\delta$ realisation, we have 14 spec-$z$ galaxies and a varying number of photo-$z$ members, with a median of 17 photo-$z$ members per realisation. To study the dependence of these physical parameters on the environment, we make a linear fit in log scale including the contribution of the error on the y-axis. This thus gives us 1000 best fit slopes and intercepts for $\tau_{\rm dep},\, \mu_{\rm gas},$ and SFR versus $\sigma_\delta$ relations. 

Once we have the best fit slopes and intercepts for all the realisations of  $\tau_{\rm dep},\, \mu_{\rm gas},$ and SFR versus $\sigma_\delta$ , we use the median slopes and intercepts of the distribution of the 1000 fits as our final best-fit relations.

\begin{figure*}
\centering
\begin{tabular}{cc}
\includegraphics[width=9cm]{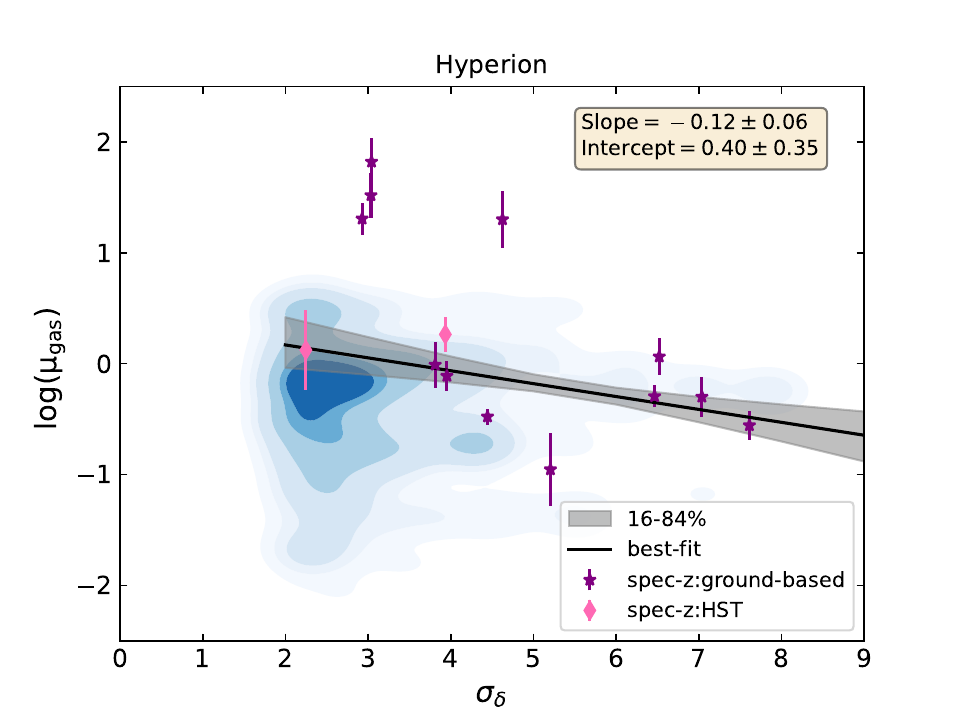}&  
\includegraphics[width=9cm]{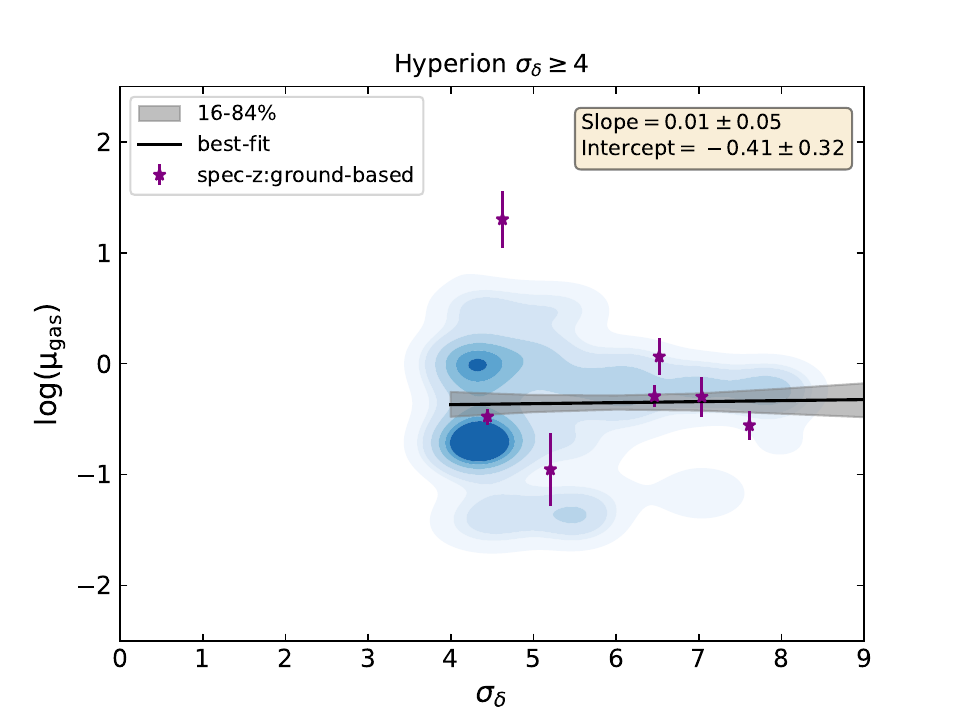} \\
\end{tabular}

\caption{\label{fig:mu_gas_sigma_slopes}Molecular gas fraction, $\mu_{\rm gas} \equiv \rm M_{\rm gas}/M_*$ as a function of the environment described in terms of $\sigma_{\rm \delta}$. The symbols and contours used in this figure are the same as Fig.\,\ref{fig:tdep_sigma_slopes}.} 

\end{figure*}


\subsection{Molecular gas depletion timescale as a function of environment}

The relation between the molecular gas depletion time and the local environment of Hyperion members is plotted in Fig.\,\ref{fig:tdep_sigma_slopes}. In both the panels, the blue contours show the distribution of the photo-$z$ galaxies combined in the 1000 realisations. The spec-$z$ Hyperion members are represented as purple stars (for ground-based spec-$z$) and pink diamonds (for \textit{HST}-Hyperion spec-$z$s). The best-fit relation is shown as the black-solid line. This is computed from the median slope and intercepts of the fits of the 1000 realisations of photo-$z$ galaxies along with the spec-$z$ galaxies (see Sect.\,\ref{sec:results}). 

In the left panel of Fig.\,\ref{fig:tdep_sigma_slopes}, we see a decrease in the gas depletion timescales at denser environments. This trend was also seen in the sample of \citet{Wang18}, where they analysed a sample of 12 spectroscopic members of Peak-5 of Hyperion. For our spec-$z$ and photo-$z$ sample in Hyperion, we see a median slope of $-0.13 \pm 0.06$ for the $\tau_{\rm dep} - \sigma_\delta$ relation.

In the right panel, we plot the $\tau_{\rm dep} - \sigma_\delta$ for Hyperion spec-$z$ and photo-$z$ member galaxies with $\sigma_\delta \geq 4$ to understand the environmental implications of depletion timescales in the densest regions of Hyperion. The relation has a similar trend compared to the trend of all Hyperion galaxies with a slope of $-0.14 \pm 0.06$. In general, we see a $\sim 2\,\sigma$ trend of decreasing $\tau_{\rm dep}$ or increasing SFE with denser environment in Hyperion, where the depletion timescale decreases by a factor of $\sim 6$ between $\sigma_\delta=2$ and $\sigma_\delta=8$.

\subsection{Molecular gas fraction as a function of environment}
Figure\,\ref{fig:mu_gas_sigma_slopes} shows the molecular gas fractions as a function of the environment in Hyperion. The contours and symbols used are similar to those in Fig.\,\ref{fig:tdep_sigma_slopes}. In the left panel of Fig.\,\ref{fig:mu_gas_sigma_slopes}, we show the evolution of $\mu_{\rm gas}$ as a function of the environment for our spec-$z$ and photo-$z$ galaxies for Hyperion. We see a decrease in the gas fraction with increasing overdensities. The best-fit slope for this relation is $-0.12 \pm 0.06$.

\begin{figure*}
\centering
\begin{tabular}{cc}
\includegraphics[width=9cm]{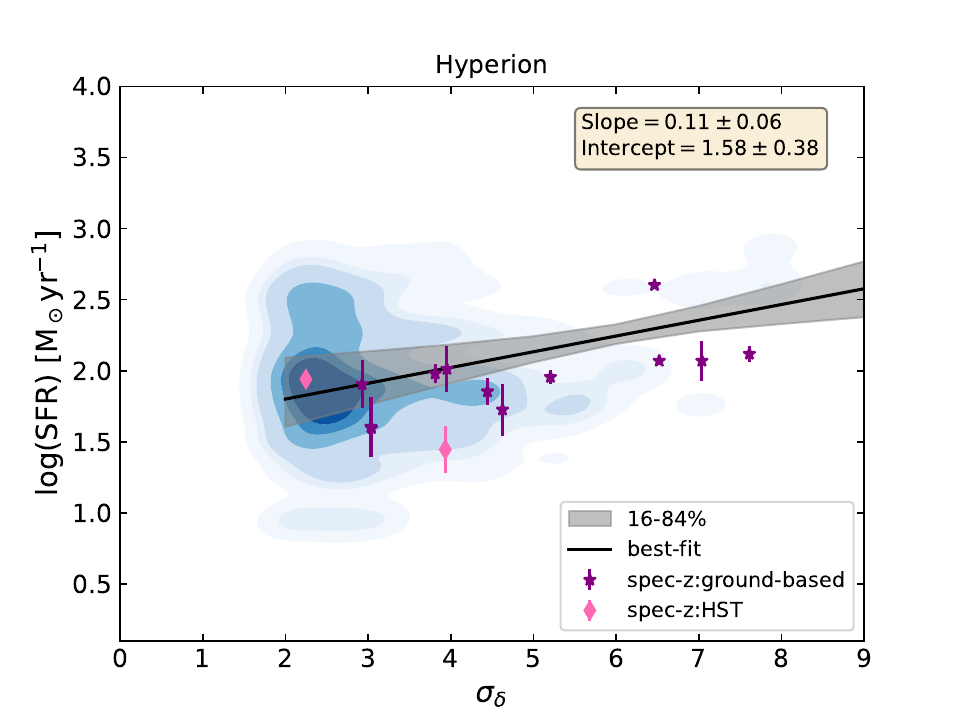}&  
\includegraphics[width=9cm]{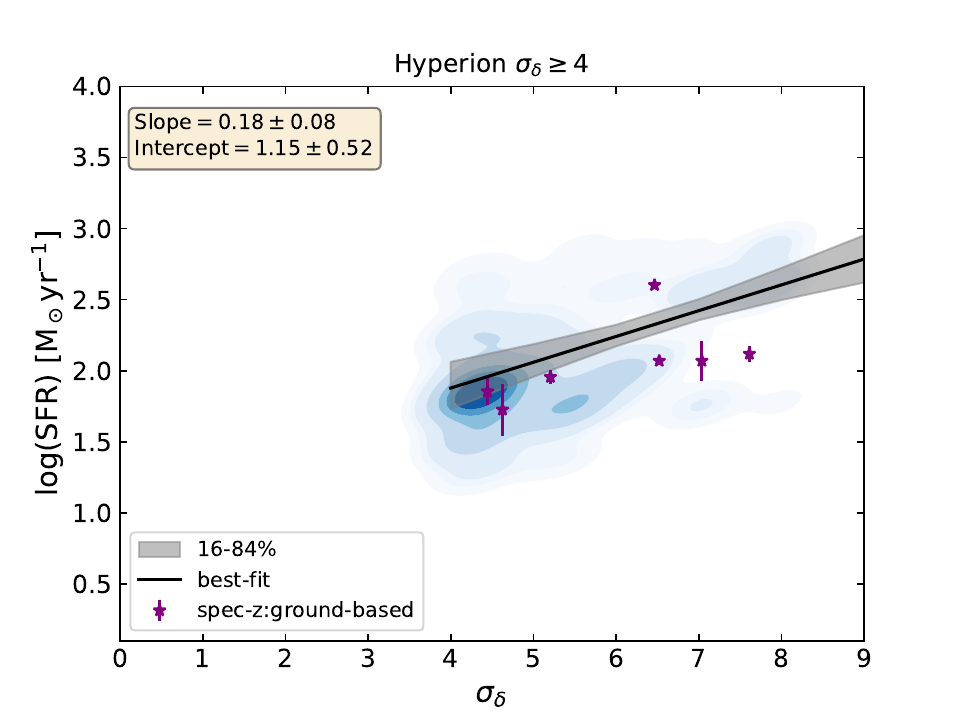} \\
\end{tabular}

\caption{\label{fig:SFR_sigma_slopes}Star-formation rate as a function of the environment ($\sigma_{\rm \delta}$). The symbols and contours used in the figure are the same as those described in Fig.\,\ref{fig:tdep_sigma_slopes}. } 

\end{figure*}


In the right panel of Fig.\,\ref{fig:mu_gas_sigma_slopes}, we compare the $\mu_{\rm gas} - \sigma_\delta$ relation for galaxies with $\sigma_\delta \geq 4$, representing the densest regions of Hyperion. Here, we see a rather flat trend in gas fractions with increasing densities, with a slope of $0.01 \pm 0.05$.

\subsection{Star-formation rates as a function of environment}
In Fig.\,\ref{fig:SFR_sigma_slopes}, we show the evolution of the star-formation rates in galaxies as a function of their environment. The contours and the symbols used are similar to those used in Fig.\,\ref{fig:tdep_sigma_slopes}. 

The SFR - $\sigma_\delta$ relation for Hyperion is shown in the left panel of Fig.\,\ref{fig:SFR_sigma_slopes}. The SFR increases with increasing densities, with a slope of $0.11 \pm 0.06$. This trend slightly strengthens in the densest regions of Hyperion ($\sigma_\delta \geq 4$, right panel of Fig.\,\ref{fig:SFR_sigma_slopes}), where we see a $\sim 2.25\sigma$ significant trend with slope of $0.18\pm0.08$ for the SFR-$\sigma_\delta$ relation in the densest regions of Hyperion.

Overall, we see an increase in SFR with decreasing depletion timescales and gas fractions within Hyperion, becoming stronger in the densest regions. These trends are similar to those reported in the literature for Hyperion (specifically, peak 5 of Hyperion presented in \citealt{Wang18}) and within other C3VO proto-clusters \citep{Lemaux22}. We further discuss the implications of these results in Sect.\,\ref{sec:discussion}.

\section{Discussions}\label{sec:discussion}

\begin{figure}[h]
\centering

\includegraphics[width=9cm]{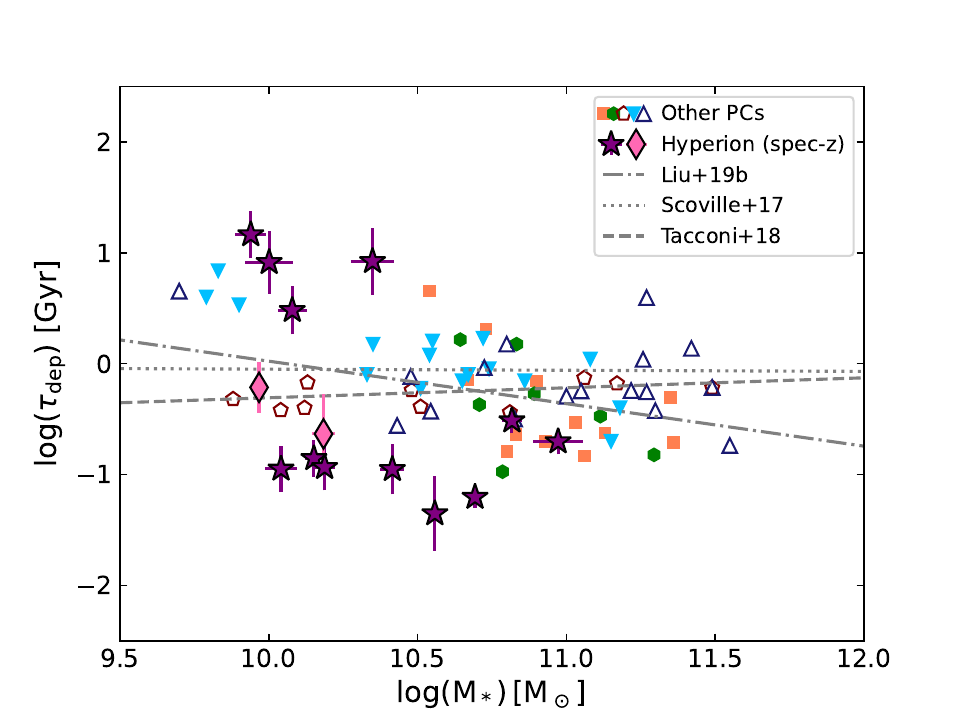}  
\includegraphics[width=9cm]{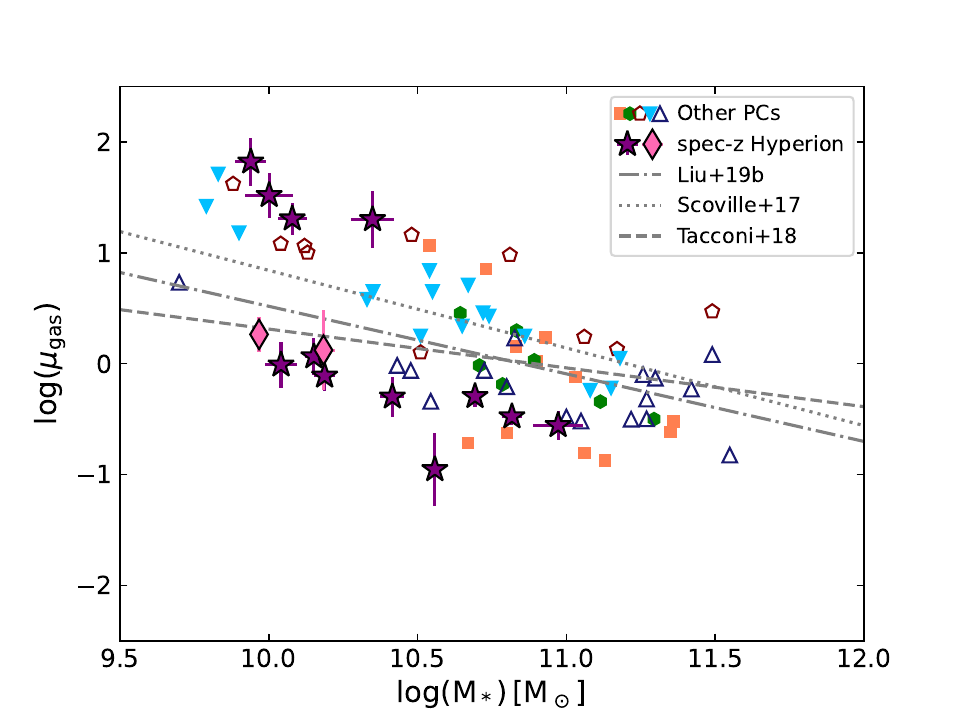} 

\caption{\label{fig:comparison to other PCs} Depletion timescale (top panel) and molecular gas fraction (bottom panel) as a function of stellar mass. The purple stars and pink diamond represent the ground-based and \textit{HST} spec-$z$ members of Hyperion respectively. The comparison sample consists of proto-cluster galaxies in the literature between $z\sim 2-3$ (\citealt{Lee17} in green-filled hexagons, \citealt{Wang18} in orange-filled squares, \citealt{GomezGuijarro19} in maroon open pentagons, \citealt{Tadaki19} in blue downward triangles, and \citealt{Zavala19} in dark-blue open triangles). The main-sequence depletion times and gas fractions at $z\sim 2.45$ from \citet{Scoville17, Tacconi18, Liu19b} are plotted as the dotted, dashed and dot-dashed lines. }
\end{figure} 

\subsection{Potential environmentally driven molecular gas depletion}

Probing the evolution of the molecular gas depletion timescales in galaxies across the various environments of Hyperion could hold clues to understanding the star-formation mechanisms and the stellar mass growth.  Within Hyperion, we see that the molecular gas depletion timescales decrease with increasing overdensities at $\sim 2.1\,\sigma$ level within Hyperion and $\sim 2.3\,\sigma$ level in the peaks of Hyperion. In other words, we find that galaxies at the densest regions in Hyperion exhibit higher star-formation efficiencies (SFE$\equiv$ SFR/M$_{\rm gas}$) than the galaxies in less-dense environments, with galaxies in denser regions appearing to have both a decreased gas reservoir and increased star-formation activity. \citet{Wang18} also found a similar trend in the sample of galaxies in the X-ray cluster CLJ1001 (corresponding to peak-5 of Hyperion), where they found higher SFEs (or shorter $\tau_{\rm dep}$) towards denser environments. 

In the case of the observed depletion timescales, we find a mean depletion timescale for the combined sample of spectroscopic and photometric members of $\sim 355 \pm 31$ Myr. This is lower than most of the main-sequence depletion times proposed for the field galaxies at $z\sim 2.45$ ($\sim$ 1.5 Gyr in \citealt{Liu19b}, $\sim$ 900 Myr in \citealt{Scoville17} and $\sim$ 570 Myr in \citealt{Tacconi18}, as shown by the three lines in Fig.\,\ref{fig:comparison to other PCs}). Four of our spectroscopic galaxies have depletion timescales larger than $\sim 3$ Gyr, longer than the field main-sequence at $z\sim 2.45$. But, these galaxies fall within the star-forming main-sequence at $z\sim 2.45$ (Fig.\,\ref{fig:MS_plots}), suggesting that their longer depletion times is likely due to the presence of a large gas reservoir rather than low star-formation rates. 

Many studies on proto-clusters have found similar depletion timescales for the member galaxies, comparable or shorter to the main-sequence depletion timescales for the field galaxies proposed in the literature. \citet{Champagne21} studied 8 spectroscopically confirmed galaxies associated to the core of the peak 1 of Hyperion and found a depletion timescale of $\sim 340$ Myr, similar to what we find within the peaks and filamentary regions of Hyperion. However, their gas masses are computed using CO emission line and the value of \aCO assumed ($6.5 \rm M_\odot/(\rm K\,km\,s^{-1}\,pc^2)$) is not comparable to the \gdr (our \gdr $=100$, gives \aCO $\sim 3 \rm M_\odot/(\rm K\,km\,s^{-1}\,pc^2)$, based on cross-calibrations between the tracers from \citealt{Dunne21, Gururajan23, Frias-Castillo24}) we assume for our sample. Similarly, \citet{Wang18} find a mean depletion timescale of $\sim 400$ Myr for the peak-5 of Hyperion. The authors have also computed the gas mass based on CO emission, assuming an \aCO $\sim 4.1 \rm M_\odot/(\rm K\,km\,s^{-1}\,pc^2)$. On both these studies, assuming a lower \aCO comparable to our \gdr, results in shorter depletion timescales than our estimated depletion timescales. Other studies on proto-clusters with CO observations find similar SFEs or $\tau_{\rm dep}$ to the main-sequence field galaxies at the same redshifts \citep{Lee17, Dannerbauer17, GomezGuijarro19, Miller18, Long20}. In the top panel of Fig.\,\ref{fig:comparison to other PCs}, we compare the depletion timescales of our spec-$z$ sample as a function of their stellar masses with other proto-cluster member galaxies in the literature \citep{Lee17,Wang18,GomezGuijarro19,Tadaki19,Zavala19} at $z\sim 2-3$. Although with a large dispersion, we find a broad agreement between the depletion scales found in Hyperion members and those found in other proto-clusters members at the same redshift (Fig.\,\ref{fig:comparison to other PCs}, top panel). However, our members mostly populate the lower stellar mass range in comparison to the other proto-clusters.

\subsection{Properties of the molecular gas reservoir as a function of the environment}

The cold molecular gas fractions of spectroscopic and photometric Hyperion member galaxies are computed based on their molecular gas masses and stellar masses. In the case of the members of Hyperion, we find a mean $\mu_{\rm gas} = 0.83 \pm 0.36$. These values are in agreement with the field main-sequence gas fractions at $z\sim 2.45$ reported by \citet{Tacconi18} ($\sim 1.2$), slightly lower than \citet{Liu19b} ($\sim 1.8$) and much lower than the values reported by \citet{Scoville17} ($\sim 7$). In the case of gas fractions in proto-clusters in comparison to the field, studies have reported similar gas fractions \citep[e.g.][]{Lee17,Zavala19,Champagne21}, lower gas fractions \citep[e.g.][]{Hill20} or higher gas fractions \citep[e.g.][]{Tadaki19, GomezGuijarro19}. In our case, we find $28\%$ our spectroscopic sample to have very high values of $\mu_{\rm gas} > 10$. Such high values in member galaxies of proto-clusters have also been reported in \citet{Tadaki19} and \citet{GomezGuijarro19}. \citet{Tadaki19} assume a higher \aCO $\sim 4.5\, \rm M_\odot/(\rm K\,km\,s^{-1}\,pc^2)$ (higher than our \gdr cross-calibrated \aCO $\sim 3$), whereas, \citet{GomezGuijarro19} assume a similar \aCO $\sim 3\, \rm M_\odot/(\rm K\,km\,s^{-1}\,pc^2)$. In the bottom panel of Fig.\,\ref{fig:comparison to other PCs}, we compare the molecular gas fractions of Hyperion members (spec-$z$ sample) to member galaxies of other proto-clusters in the literature \citep{Lee17,Wang18,GomezGuijarro19,Tadaki19,Zavala19} at $z\sim 2-3$. Overall, we find a broad agreement with the other proto-cluster member galaxies.

The evolution of molecular gas fractions as a function of the environment is shown in Fig.\,\ref{fig:mu_gas_sigma_slopes}. In the left panel, we see a $2\,\sigma$ trend of decreasing gas fractions at denser environments within Hyperion. Nonetheless, this trend becomes rather flat when comparing only the densest regions of Hyperion. This might indicate a transition from gas-rich members in the rarer-filamentary regions of Hyperion which become gas poor while approaching the denser regions. \citet{Wang18} find similar trends for the evolution of gas fractions with proximity to the cluster center (a proxy to the denser regions). They find a sharp transition between gas rich members in the outskirts to gas poor members in the center. The spatial scales considered in \citet{Wang18} are roughly equivalent to the Hyperion $\sigma_\delta \geq 4$ regions.

\subsection{Star formation as a function of the environment in Hyperion}

Evolution of the star-formation rates of galaxies as a function of the environment has been probed for many proto-clusters and low-$z$ clusters \citep[e.g.][]{Capak11, Hatch11,
Wang16, Lemaux22, Staab24}. While in low- and intermediate-$z$ ($z\la1.5$) clusters, the SFR tends to decrease towards the cluster center hosting more massive and passive galaxies \citep[e.g.][]{Tomczak19, Lemaux19, Old20}, \citet{Lemaux22} found that, in high-$z$ proto-clusters, SFRs increase at denser environments (see also \citealt{Alberts14,Shimakawa18,Monson21,Laishram24,Taamoli24}). This reversal in the SFR-environment trend at $z>2$ indicates that, at this epoch, cluster-center galaxies could be more active and star-forming, and would transition into more passive systems in the future. In our case, we also find a positive correlation between the SFR and the environment among Hyperion members, with more intensely star-forming systems found in the denser environments ($\sim 1.8\,\sigma$ within Hyperion and $\sim 2.25\,\sigma$ in the peaks of Hyperion). These systems also have a higher star-formation efficiency (Fig.\,\ref{fig:tdep_sigma_slopes}, shorter depletion timescales), but do not show signs of enhanced gas fractions. In contrast, we observe a decrease in gas fractions as we move from less dense, filamentary regions ( $\sigma_\delta \lesssim 4$) to the densest environments in Hyperion (Fig.\,\ref{fig:mu_gas_sigma_slopes}, left panel). This increase in SFR at denser environments despite no significant increase in gas fractions could indicate that mechanisms other than inflowing gas could be responsible for driving the SFRs in the densest regions of Hyperion. Furthermore, physical processes such as merging events, tidal torquing or gas compression, ram-pressure stripping could further impact the SFR trends seen in these galaxies \citep[e.g.][]{Boselli14, Bialas15, Coogan18}. 

\subsection{Environmentally-driven galaxy evolution in Hyperion}

In our analysis, we see that galaxies in the densest regions of Hyperion show enhanced star-formation rates and efficiencies while having a deficit of molecular gas with respect to galaxies inhabiting less dense environments. However, the trends that we measure are at a significance of $\sim 2\, \sigma$, and, thus, we need larger, uniformly-selected samples across various environments to be more conclusive on the environmental impact on these galaxy properties. From Figs.\,\ref{fig:tdep_sigma_slopes},\,\ref{fig:mu_gas_sigma_slopes}, and \ref{fig:SFR_sigma_slopes}, we find an increase in the SFR, increase in the SFE (decreasing depletion timescale) and flat/slight decrease in gas fractions with increasing overdensities within Hyperion. Thus, the increased efficiencies are mainly driven by the increased SFR. The suppression in molecular gas seen within Hyperion can be attributed to many possible reasons such as compaction of gas due to tidal forces \citep[e.g.][]{Mok17}, ram-pressure stripping \citep[e.g.][]{Cortese10,Wetzel13,Bahe15,Matharu21} or the lack of gas accretion from the cosmic-web \citep[e.g.][]{Dekel06,Dekel09}. 

We also see a tentative, $\sim 2\sigma$ correlation between the depletion times and stellar mass for the spectroscopic galaxies within Hyperion (Fig.\,\ref{fig:Stellar mass sigma relation}, Spearman correlation coefficient $r = -0.52$ and $p-$value $=0.06$). Galaxies at the high-mass end (e.g. log(M$_*/\rm M_\odot) \sim 10.25 - 10.5$) of the stellar mass function likely have to start transitioning en masse to reproduce the high quenched fractions seen for such masses at $z\sim 1-1.5$ \citep[e.g.][]{Nantais17,Lemaux19, Baxter23}. This gives $\sim 1.5-2.5$ Gyr until $z\sim 1.5-1$ for the galaxies to be quenched, and it takes around $\sim 1$ Gyr to transition into the red-sequence once the star-formation has ceased. The mass range log(M$_*/\rm M_\odot) \sim 10.25 - 10.5$ is also the preferential mass that is seen to begin transitioning to quiescence in $z\sim1$ clusters \citep[e.g.][]{Tomczak19}. The tentative trends we observe in our $\tau_{\rm dep},\, \mu_{\rm gas}$ and SFR further supports this possible transition, as the galaxies in the central regions appear to exhaust their molecular gas supply in $\sim 350$ Myr (assuming no further gas accretion) after which they would begin transitioning into quiescence. Furthermore, at $z\sim 2-3$, the cold gas inflows in massive halos (similar to the peaks of Hyperion) are likely to be cut-off due to virial shock heating \citep[e.g.][]{Overzier16}, which further supports the transition into quiescence.

For studies on gas properties using dust as a proxy, there could be potential caveats induced by the assumption of the gas-to-dust ratio, \gdr, primarily due to its dependence on metallicities (decreasing \gdr at increasing metallicities). In the case of solar metallicities, our assumed value of \gdr $=100$ would be well suitable \citep[e.g.][]{Magdis12, RemyR14}, but, we currently do not have an estimate of the metallicities in the member-galaxies of Hyperion, although, could potentially be constrained in the near-future with the availability of [OIII], [OII] and H$_\beta$ lines from C3VO-Mosfire and \textit{HST-}Hyperion. The environmental dependence of the metallicities of galaxies specifically in high-$z$ ($z\ga$ 2) proto-clusters has been probed in the literature, but remains poorly constrained. In the case of low-$z$ clusters, most studies find that the metallicities increase in dense environments \citep[e.g.][]{Cooper08, Ellison09, Peng14, Wu17, Schaefer19}. In the case of proto-clusters at $z\ga 2$, the environmental dependence on metallicity is debated, some studies find a decreasing metallicity at dense environments \citep[e.g.][]{Valentino15, Chartab21,Sattari21}, while others find no environmental implications \citep[e.g.][]{Kacprzak15, Tran15, Namiki19} or an increasing metallicity at dense environments \citep[e.g.][]{Kulas13,Shimakawa15,Perez-Martinez23}.

\section{Summary}\label{sec:conclusions}

In this paper, we presented a sample of spectroscopically confirmed member galaxies with detections in the FIR photometric bands of the proto-supercluster Hyperion along with potential members with only photometric redshifts. Through this study, we probe  the gas reservoirs in the galaxies in Hyperion as a function of the local environment they reside in. 

The proto-supercluster Hyperion is extremely large and massive, containing seven overdense peaks connected by less-dense, filamentary regions. The photometric observations for these galaxies are primarily derived from the C20-Farmer catalogue, along with the COSMOS-super-deblended catalogue in the FIR wavelengths and the A$^3$COSMOS catalogue in the sub-millimeter. The spectroscopic confirmation of our sample mainly comes from the C3VO catalogue that contains spectroscopic observations using MOSFIRE and DEIMOS on Keck and the \textit{HST}-Hyperion programme along with the ancillary spectroscopic measurements from VUDS, zCOSMOS, DEIMOS10k and MAGAZ3NE surveys. The environmental properties of these galaxies are obtained from the density field maps constructed by \citet{Cucciati18} using the VMC mapping technique. We select a sample of galaxies with detections in the FIR/sub-mm wavelengths and our final sample consists of 14-spectroscopically confirmed Hyperion member galaxies (spec-$z$ sample) and 616 potential Hyperion member galaxies containing high-quality photometric redshifts, without a spectroscopic redshift (photo-$z$ sample). 

The stellar masses, SFRs and gas masses of our sample are constrained by fitting the SEDs of the galaxies between optical and radio wavelengths (for SFR and stellar mass) and FIR wavelengths (for the gas mass) using \texttt{CIGALE} and a simple MBB for the dust emission. From the derived physical properties and the environmental properties, we explore the impact of the environment on the gas content in the galaxies of Hyperion. Below we highlight the main results of this study: 

\begin{itemize}
    \item Although the molecular gas depletion timescales of the member-galaxies of Hyperion are in broad agreement with the values reported in other proto-cluster at $z\sim 2-3$, our sample hosts shorter depletion times than the field galaxies at $z\sim2.45$. 
    \item We observe a decrease in the depletion timescale with increasing overdensities, both within Hyperion ($\sim 2.1\sigma$) and specifically in the densest regions of Hyperion ($\sim 2.3\sigma$) . This is similar to the results obtained in \citet{Wang18}, where they report an increased efficiency (decreased depletion time) in the cluster-center (densest regions of the proto-supercluster). 
    \item While studying the fraction of molecular gas, we find similar values to those observed in the field, though, with a large scatter. Additionally, 4/14 of our spectroscopically confirmed galaxies show the presence of a large gas fraction in them similar to those reported in \citet{Tadaki19} and \citet{GomezGuijarro19}. 
    \item The gas fractions of member-galaxies in Hyperion decreases with increasing overdensities (we observe a $\sim 2\sigma$ trend), similar to those reported by \citet{Wang18} but, this trend becomes rather flat in the densest regions of Hyperion. This suggests that the galaxies in the filamentary regions of Hyperion could be more gas rich than those in the peaks of Hyperion. 
    \item We also see a slight trend between the SFR and the environment in the member galaxies of Hyperion ($\sim 1.8\sigma$), similar to those reported for high-$z$ proto-clusters \citep[e.g.][]{Lemaux22}. The SFR in the peaks of Hyperion increases more steeply with increasing overdensities ($\sim 2.25\sigma$ trend). 
    \item Thus, galaxies in the densest environments of Hyperion show an increase in the SFE, which is mainly supported by an increasing SFR and reinforced by a decreasing molecular gas fractions in the denser regions.  
\end{itemize}

In this work, we use a sample of 14 galaxies with robust spectroscopic redshifts available and in the stellar mass range of log(M$_*$/M$_\odot$) $\sim 9.8 - 11.3$ along with 616 galaxies with robust photometric redshifts. However, deriving the precise environmental properties associated to each galaxy requires a secure spectroscopic redshift. Even though our results hint towards ($\sim 2\sigma$) dependencies on the environment in the properties of the gas reservoirs in galaxies, we need larger, uniformly selected samples to obtain a concrete understanding of environmental implications in gas reservoirs of galaxies. Thus, dedicated follow-ups in the sub-mm could not only help increase the spectroscopic confirmation of Hyperion member galaxies, but also help constrain their gas properties. Additionally, testing these relations on a large number of uniformly selected proto-clusters could help in providing a global overview on the impact of the environment and probe any structure-to-structure variations.

\begin{acknowledgements}
We thank the anonymous referee for their valuable comments. GG acknowledges support  from the grants PRIN MIUR 2017 - 20173ML3WW$\textunderscore$001, ASI n.I/023/12/0 and INAF-PRIN 1.05.01.85.08. GG, OC, RD, MT, DV, and, SB acknowledge support from the INAF mini-grant 2023 "LION: Looking for the Imprint of Overdensity Networks". DCB is supported by an NSF Astronomy and Astrophysics Postdoctoral Fellowship under award AST-2303800. DCB is also supported by the UC Chancellor's Postdoctoral Fellowship. Some of this work is based on observations taken by HST Program 16684 with the NASA/ESA HST, which is operated by the Association of Universities for Research in Astronomy, Inc., under NASA contract NAS5-26555. The authors wish to recognise and acknowledge the very significant cultural role and reverence that the summit of Maunakea has always had within the indigenous Hawaiian community. We are most fortunate to have the opportunity to conduct observations from this mountain. This work is also based on observations collected at the European Southern Observatory under ESO programmes 175.A-0839, 179.A-2005, and 185.A-0791, as well as work supported by the National Science Foundation under Grant No. 1908422. Some of the data presented herein were obtained at Keck Observatory, which is a private 501(c)3 non-profit organization operated as a scientific partnership among the California Institute of Technology, the University of California, and the National Aeronautics and Space Administration. The Observatory was made possible by the generous financial support of the W. M. Keck Foundation. This work was supported by NASA’s Astrophysics Data Analysis Program under grant number 80NSSC21K0986. This work was also supported by the international Gemini Observatory, a program of NSF NOIRLab, which is managed by the Association of Universities for Research in Astronomy (AURA) under a cooperative agreement with the U.S. National Science Foundation, on behalf of the Gemini partnership of Argentina, Brazil, Canada, Chile, the Republic of Korea, and the United States of America. Some of the data used in this work is based on observations collected at the European Southern Observatory under
ESO programme ID 179.A-2005 and on data products produced by CALET and
the Cambridge Astronomy Survey Unit on behalf of the UltraVISTA consortium.

\end{acknowledgements}
\bibliographystyle{aa} 
\bibliography{bibliography.bib}
\begin{appendix}

\section{Star-forming main-sequence in spec-$z$ Hyperion member galaxies}

\begin{figure}
\centering

\includegraphics[width=9cm]{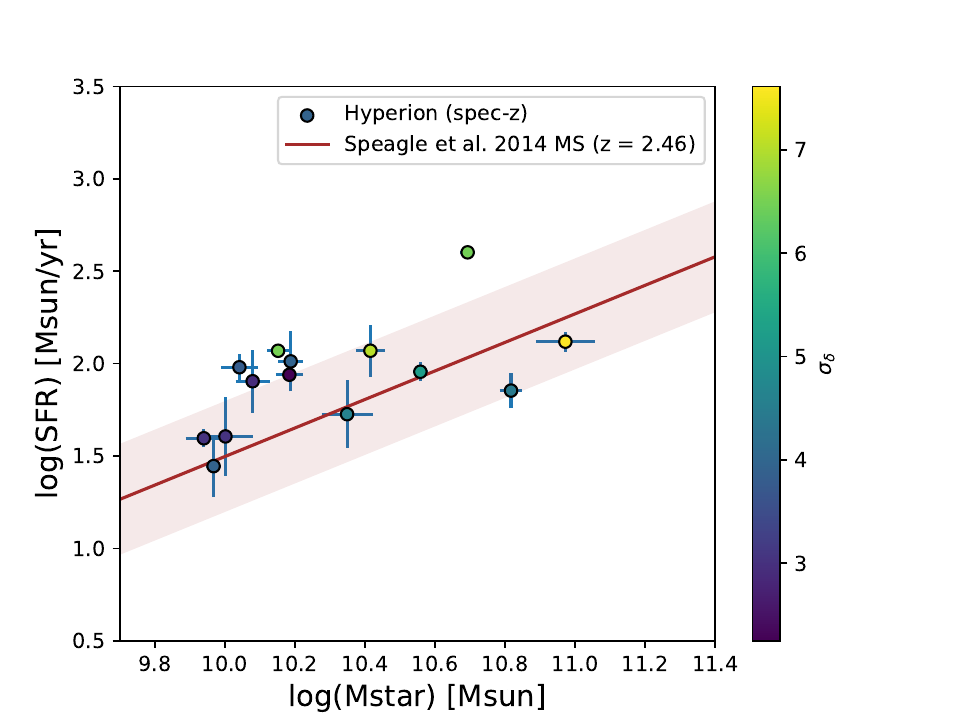} \\

\caption{\label{fig:MS_plots} Star-formation rate versus stellar mass plot for our sample of spec-$z$ galaxies. The star-formation rate versus stellar mass of the Hyperion member galaxies with a spectroscopic redshift. The overdensity $\sigma_\delta$ associated with these galaxies are shown in the colorbar. The star-forming main-sequence relation at $z=2.46$ from \citet{Speagle14} is shown in by the red solid line.  } 

\end{figure}

The star-formation rate versus stellar mass of the Hyperion member galaxies with a spectroscopic redshift are plotted in Fig.\,\ref{fig:MS_plots}. We see that majority of the spec-$z$ Hyperion members are on the star-forming main-sequence at the given redshift. Thus, the longer depletion timescales seen in some of the spec-$z$ members in Fig.\,\ref{fig:tdep_sigma_slopes} can be attributed to the presence of larger gas reservoirs in galaxies that host main-sequence like SFRs.

\section{Evolution of stellar mass with the environment for FIR-selected Hyperion galaxies}

\begin{figure}
\includegraphics[width=9cm]{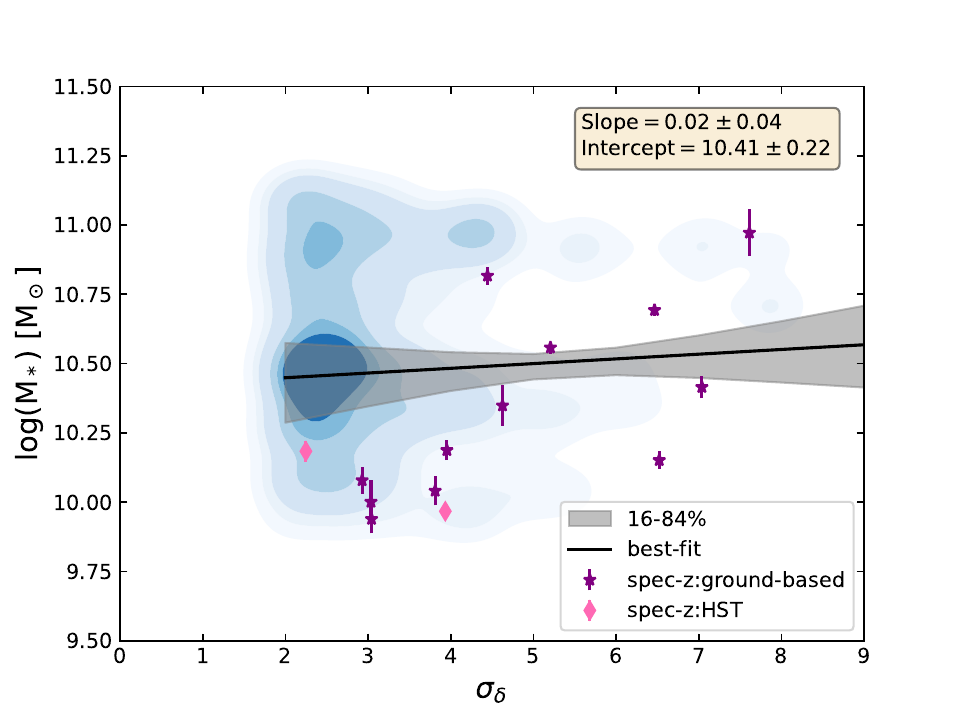}
\caption{\label{fig:Stellar mass sigma relation} Stellar mass as a function of the local overdensity of galaxies in Hyperion. The blue-contours represent the distribution of the photo-$z$ galaxies from their 1000 realisations. The purple stars are the spec-$z$ Hyperion members, the pink diamonds represent the spec-$z$ galaxies from \textit{HST}-Hyperion sample. The best fit relation and the 1-$\sigma$ error are shown as the black-solid line and the grey shaded regions respectively. The best-fit slopes and intercepts are written in the inset-yellow panel. The errors on the slopes and intercepts are the standard deviation of the slopes and intercepts of the 1000 relations. } 
\end{figure}

The relation between the stellar-mass and the overdensity $\sigma_\delta$ of the spec-$z$ and photo-$z$ Hyperion member galaxies are shown in Fig.\,\ref{fig:Stellar mass sigma relation}. The stellar mass of our sample of galaxies do not show a strong dependence on their local overdensities. Thus, this indicates that the stellar mass does not play a major role in driving the trends of depletion timescales, molecular gas fractions and SFRs as a function of the overdensity (see Figs.\,\ref{fig:tdep_sigma_slopes},\,\ref{fig:mu_gas_sigma_slopes}, and \ref{fig:SFR_sigma_slopes}). 

\end{appendix}

\end{document}